\documentclass[aps,prl,reprint,longbibliography]{revtex4-1}
\usepackage{graphicx}
\usepackage{hyperref}
\usepackage{physics}

\usepackage{amsmath}
\usepackage{color}

\usepackage{amssymb}
\usepackage{amsfonts}
\usepackage{bm}
\usepackage[T1]{fontenc}
\usepackage[utf8]{inputenc}

\newcommand{\QA}{Q_-}
\newcommand{\QB}{Q_+}
\newcommand{\QS}{Q_{\Sigma}}
\newcommand{\QD}{Q_{\Delta}}

\newcommand{\myhat}[1]{#1}

\begin{document}

\title[]{Atom-only theories for U(1) symmetric cavity-QED models}

\author{Roberta Palacino$^1$, Jonathan Keeling$^{1}$} 

\address{$^1$ SUPA, School of Physics and Astronomy, University of St Andrews, St Andrews, KY16 9SS, United Kingdom}


\begin{abstract}
We consider a generalized Dicke model with U(1) symmetry, which can undergo a transition to a superradiant state that spontaneously breaks this symmetry. By exploiting the difference in timescale between atomic and cavity dynamics, one may eliminate the cavity dynamics, providing an atom-only theory. We show that the standard Redfield theory cannot describe the transition to the superradiant state, but including higher-order corrections does recover the  transition.  Our work reveals how the forms of effective theories must vary for models with continuous symmetry, and provides a template to develop effective theories of more complex models.
\end{abstract}

\maketitle

\textit{Introduction}---  While phase transitions in equilibrium many-body systems have been extensively studied and are broadly understood~\cite{landau1980statistical, chaikin1995principles}, the critical behavior of driven-dissipative out-of-equilibrium systems~\cite{Diehl2008:quantum,Carusotto2013:RMP,LeHur2016:Many,Chang2018:RMP} poses  open questions.  
A central challenge in numerical exploration of such systems is the exponential growth of Hilbert space dimension with problem size.   As such, any ability to reduce Hilbert space size, e.g. by \textit{a priori} identifying a low-energy (slow) subspace can be very powerful~\cite{Kessler2012:generalized,Sciolla2015:TwoTime,Lenarcic2018:Perturbative,Saideh2020:Projection}.  A widely-used approach to derive a reduced model is  Redfield theory~\cite{Redfield1957a, Breuer2002}. Previous work~\cite{Damanet2019:AtomOnly} has shown this works well for the steady states and collective modes of the Dicke model. However, as we show below, Redfield theory can fail for models with U(1) symmetry, requiring higher-order approaches~\cite{Muller2017:Deriving}.

A leading platform to study driven-dissipative many-body physics is cold atoms~\cite{Bloch:2008td} in dissipative optical cavities~\cite{Ritsch03_review}, using Raman driving~\cite{Dimer2007:Proposed}. This platform has been studied in a wide variety of experiments~\cite{Baumann2010,Hemmerich14,Klinder2015,Landig2016,Kollar2017sm,Leonard2017:Supersolid,Leonard2017:Supersolid,vaidya2018tunable,Landini2018:Formation,Kroeze2018:Spinor,Zhang2018:Dicke,Guo2019Sign,Kroeze2019:Dynamical,Davis2019:Photon}.
In its simplest form, an ensemble of $N$ atoms coupled to a cavity via Raman pumping realizes a Dicke model~\cite{Dimer2007:Proposed}. Such a model has a critical pumping strength, above which it undergoes a phase transition to a superradiant state, spontaneously breaking a  $\mathbb{Z}_2$ symmetry~\cite{hepp1973equilibrium,wang73,Garraway2011,kirton2019:Review}.  One can also engineer more complex models, with multiple photon modes, which have U(1)~\cite{Fan2014:Hidden,baksic2014controlling, Leonard2017:Supersolid,Leonard2017:Monitoring,Moodie2018:Generalized} or higher symmetries~\cite{Chiacchio2018:Emergence}.
Furthermore, by using a degenerate (e.g. confocal) cavity, one can also explore symmetry breaking in spatially extended  systems~\cite{Gopalakrishnan09,Gopalakrishnan10,Labeyrie:2014gh}, as well as engineering exotic light-matter phases \cite{Kollar2017sm, ballantine2017meissner,Kroeze2018:Spinor,Kroeze2019:Dynamical,rylands2020photon}, tunable atomic interactions \cite{vaidya2018tunable, Guo2019Sign,Guo2019Emergent}, and quantum memories \cite{Gopalakrishnan2012:exploring,Fiorelli2020:Signatures,marsh2020enhancing}. In contrast to single- or few-mode systems, continuous symmetry breaking in extended multimode systems allows one to explore the dispersion of the (complex) Goldstone modes~\cite{Carusotto2013:RMP,Minganti2018:Spectral} and their contribution to critical behavior.
As noted above, theoretically modeling such systems is challenging; one fruitful approach is to adiabatically eliminate fast degrees of freedom, providing an effective theory of the slow and gapless degrees of freedom.  This idea motives the current work.

In this Letter, we consider a two-mode generalized Dicke model which has U(1) symmetry~\cite{Moodie2018:Generalized}, and discuss how an effective theory can be developed.   Given the motivation above,  there are certain conditions required of an effective theory: it must describe the transition to a symmetry broken state, and must correctly describe the frequencies and damping rates of the low-energy (soft modes) associated with this symmetry breaking. In the following, we first show why standard Redfield theory fails, and then present an alternative method which succeeds.   We also discuss how, for such an effective model, we can derive semiclassical equations of motion, applicable in the large $N$ limit.


\begin{figure}
\begin{centering}
\hspace*{-.3cm}\includegraphics[scale=.88]{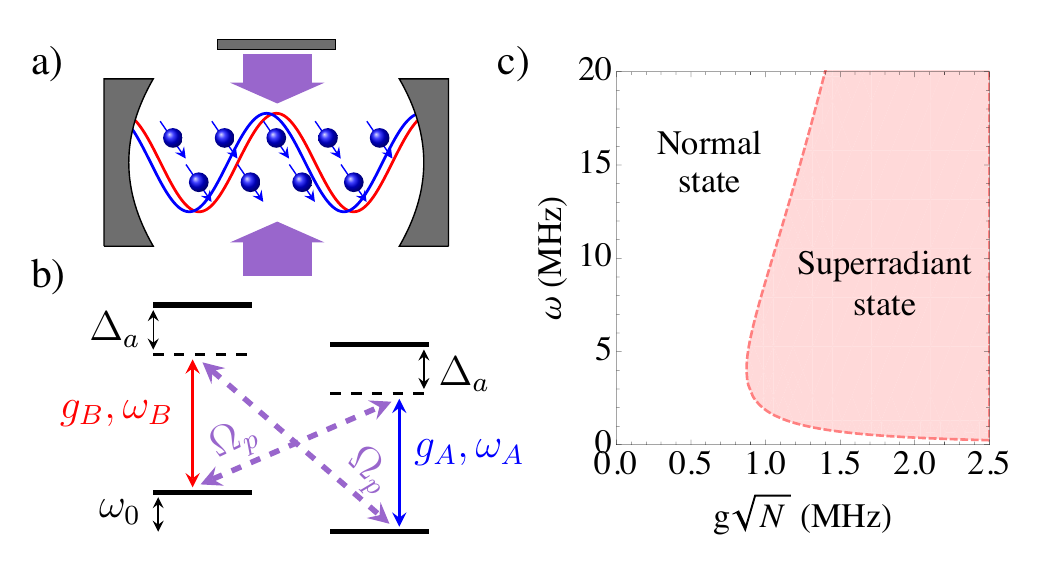}
 \par\end{centering}
  \caption{a) Experimental setup: Atoms are coupled to cavity modes (red and blue) and pumped transverse to the cavity (amethyst) b) Raman driving scheme. c) Phases of the system. The pink dashed line represents the critical coupling $g_{c}\sqrt{N}$. Parameters: $\kappa=8.1$ MHz, $\omega_{0}=47$ kHz, $\omega_{A}=\omega_{B}=\omega$.}
\label{figure1}
\end{figure}

\textit{U(1) symmetric model}--- 
We begin by introducing the $U(1)$ symmetric model that we will consider, and summarizing its mean-field behavior.
The model, introduced in Ref.~\cite{Moodie2018:Generalized}, describes N two-level systems---described via a collective spin degree of freedom---interacting with two cavity modes.
As shown in Fig.~\ref{figure1}(a,b), this can be realized by a Raman driving scheme which couples two low-energy atomic states~\cite{Dimer2007:Proposed}.  As shown, two transition pathways exist. Each pathway involves a different cavity mode---this is key to realizing a $U(1)$ symmetry.  
Including cavity losses, we find an equation of motion for 
density matrix of the total system, $\rho_t$:
\begin{align}
\dot{\myhat{\rho}}_{t}&=-i\left[\myhat{H},\myhat{\rho}_{t}\right]+\frac{\kappa}{2}\left(\mathcal L\left[\myhat{a}\right]+\mathcal L\left[\myhat{b}\right]\right),
\label{ME}
\\
\myhat{H}&=\omega_{A}\myhat{a}^{\dagger}\myhat{a}+\omega_{B}\myhat{b}^{\dagger}\myhat{b}+\omega_{0}\myhat{S}^{z}\!+g\!\left[\!\left(\myhat{a}^{\dagger}\!+\myhat{b}\right)\!\myhat{S}^{-}\!\!+\text{H.c.}\right]\!.
\label{genH}
\end{align}
Here $\myhat{a},\myhat{b}$ are the two cavity mode annihilation operators, while $\myhat{\mathbf{S}}$ is a collective spin of the atoms, with modulus $N/2$.  Cavity loss at rate $\kappa$ is described by Lindblad terms $\mathcal{L}[\myhat{X}]=2\myhat{X}\myhat{\rho}\myhat{X}^{\dagger}-\{\myhat{X}^{\dagger}\myhat{X},\myhat{\rho}\}$.  In the Hamiltonian, the energies $\omega_{A,B}$ describe the cost of scattering a photon from the pump into each cavity mode, while $\omega_{0}$ is the
splitting of the two hyperfine atomic levels. The effective coupling between light and matter is given by $g=g_{0}\Omega_{p}/\Delta_{a}^2$, where $g_{0}$ is the bare atom-cavity coupling, $\Omega_{p}$ the Rabi frequency of the pump field, and $\Delta_{a}$ is the detuning between the pump and the atomic resonance, see Fig.~\ref{figure1}(b). As shown in Ref.~\cite{Moodie2018:Generalized}, this model is invariant under a transformation $\myhat\rho \to \myhat U \myhat \rho \myhat U^\dagger$, with $\myhat U=\exp[i\theta (S^z + \myhat a^\dagger \myhat a - \myhat b^\dagger \myhat b)]$, which transforms $(\myhat{a},\myhat{b},\myhat{S}^{\pm})\to(\myhat{a}e^{i\theta},\myhat{b}e^{-i\theta},\myhat{S}^{\pm}e^{\mp i\theta})$. This corresponds to a $U(1)$ symmetry.  

For large $N$, the composite atom-cavity system is well described by semiclassical equations~\cite{Kirton2017:Suppressing}  for $\langle{\myhat a}\rangle, \langle{\myhat b}\rangle, \langle{\myhat{\vb{S}}}\rangle$, which show two distinct types of steady state. At small $g$, there is a normal phase with $\langle \myhat S^{\pm}\rangle=\langle \myhat a \rangle=\langle \myhat{b}\rangle=0$,  respecting the $U(1)$ symmetry.  At large $g$, there is a superradiant phase with a non-zero photonic field, $\langle \myhat a \rangle, \langle \myhat b \rangle \neq 0$.  Thus,
at a critical value $g=g_c$---which, for $\omega_A=\omega_B=\omega$ obeys $2 g_c^2 N=  \omega_0 (\omega^2 + \kappa^2/4)/\omega$---the system  undergoes a continuous phase transition to a state which spontaneously breaks the U(1) symmetry as shown in Fig.~\ref{figure1}(c).  Because of the cavity loss, these steady states are attractors of the dynamics,  and the system undergoes damped relaxation towards these states.

In typical experiments, there is a separation of timescales between the atomic and cavity degrees of freedom, $\kappa \gg \omega_0, g\sqrt{N}$. We thus next consider how to adiabatically eliminate the cavity degrees of freedom and still describe the same behavior as discussed above.

\textit{Adiabatic elimination at semiclassical level}---
We first consider adiabatic elimination of $\langle \myhat a \rangle, \langle \myhat b \rangle$ from the semiclassical equations given in~\cite{Moodie2018:Generalized}.  This yields an equation of motion for $\vb{S}=\langle \myhat{ \vb{S}}\rangle$. The resulting equation is conservative, defined by a Poisson bracket $\dot{\vb{S}} = \{ \vb{S}, H_{\text{sc}}\}$ where, for $\omega_A=\omega_B=\omega$, the classical Hamiltonian takes the form $H_{\text{sc}}=\omega_{0}S^{z}+\frac{2g^{2}\omega}{\omega^2 + \kappa^{2}/{4}}(S^{z})^{2}$.  While this Hamiltonian has a ground state phase transition at $g=g_c$ as expected, the purely conservative dynamics is
in contrast with the dissipative evolution expected for an open system.
Similar behavior was found for the single-mode Dicke model~\cite{Keeling2010:Collective,Damanet2019:AtomOnly}. The results of~\cite{Damanet2019:AtomOnly} suggest that to recover the correct dissipative dynamics one should instead eliminate the cavity degrees of freedom in a quantum model and then derive the semiclassical limit. 


\textit{Redfield Theory}---  To eliminate the cavity modes in a full quantum model, we use the standard Redfield approach~\cite{Redfield1957a,Breuer2002}.  Specifically, we take the collective spin as the system, and all other modes form the bath.  The system-bath coupling in the interaction picture is $\myhat{H}_{I}=g[\myhat{S}^{+}(t)\myhat{X}(t)+\myhat{S}^{-}(t)\myhat{X}^{\dagger}(t)]$, where  $\myhat{S}^{\pm}(t)=\myhat{S}^{\pm}e^{\pm i\omega_{0}t}$ and $\myhat{X}(t)=\myhat{a}(t)+\myhat{b}^{\dagger}(t)$. The time evolution of $\myhat{X}(t)$ is discussed below.  Redfield theory states:
\begin{equation}
\dot{\rho}=-\int_{-\infty}^{t}dt^{\prime}\text{Tr}_{B}\bigl(\left[H_{I}(t),\left[H_{I}(t^{\prime}),\rho\left(t\right)\right]\right]\bigr).
\label{redfield}
\end{equation}
Evaluating this requires  two-time correlations of the bath operators $\myhat X(t)$.
These are found by solving Heisenberg-Langevin equations for the cavity modes, including loss, giving
$\langle \myhat{X}(t)\myhat{X}^{\dagger}(t^{\prime}) \rangle=e^{-i\omega_{A}|t-t^{\prime}|-\frac{\kappa}{2}|t-t^{\prime}|}$,
$\langle \myhat{X}^\dagger(t)\myhat{X}^{}(t^{\prime}) \rangle=e^{-i\omega_{B}|t-t^{\prime}|-\frac{\kappa}{2}|t-t^{\prime}|}$.
In the Schr\"odinger picture, the $2^{\text{nd}}$-order Redfield  equation (2RE) is thus:
\begin{multline}
\dot{\rho}=-i\omega_{0}\left[\myhat{S}^{z},\rho\right] 
-2g^{2}\left[\QA\left(\myhat{S}^{+}\myhat{S}^{-}\rho-\myhat{S}^{-}\rho\myhat{S}^{+}\right)+\right.\\
\left.\QB\left(\myhat{S}^{-}\myhat{S}^{+}\rho-\myhat{S}^{+}\rho\myhat{S}^{-}\right)+ \text{H.c.}\right]
\label{gamma0}
\end{multline}
where $Q_{\mp}=\left[\kappa+2i\left(\omega_{A(B)}\mp\omega_{0}\right)\right]^{-1}$.

%

While this equation includes dissipative effects, as we discuss next, it does not show a phase transition with increasing $g$.  To understand why, we first rewrite this equation in Linblad form $\dot{\rho} = -i[H_{\text{2RE}},\rho] + 4 g^2 (\Re[\QA] \mathcal{L}[S^-] + \Re[\QB] \mathcal{L}[S^+])$, where  $H_{\text{2RE}}=\omega_0 S^z+2g^2(\Im[\QA]S^+ S^-+\Im[\QB]S^- S^+)$ commutes with $\myhat{S}^z$. The U(1) symmetry means the steady state must commute with $S^z$, which implies $\rho = \sum_M P_M \ket{M}\bra{M}$, where $\myhat S^z \ket{M}=M\ket{M}$.  We find $P_M$ obeys $P_{M}/P_{M+1}=\Re[\QA]/\Re[\QB]$. Since this ratio is independent of $g$,  no transition occurs at $g=g_c$.  At large $N$ we find $\langle S^z \rangle = - N/2$ for $\Re[\QA]>\Re[\QB]$, or $\langle S^z \rangle=+N/2$ if $\Re[\QA]<\Re[\QB]$---i.e., the system is always in a normal or inverted state. 
This absence of a phase transition depends on two features of the equation.
First, symmetry ensured that both the effective Hamiltonian and steady-state density matrix are diagonal in the $\myhat{S}^z$ basis, so they commute. This means the steady state depends only on the dissipative terms.  Such a statement will always be true for U(1) symmetric models.  Secondly, the dissipative terms in the Redfield equation are all proportional to $g^2$, so no $g$ dependence occurs in their ratio.  This statement is not generic, so we next consider how contributions of higher-order in $g$ change the equation.

\textit{$4^{\text{th}}$-order Keldysh--Redfield Theory}---  A systematic method to derive higher-order density matrix equations was introduced by \citet{Muller2017:Deriving}, making use of Keldysh diagrammatic perturbation theory. 
This technique allows one to take into account all  contributions at each order while avoiding double counting. We write the density matrix equation in the form $\dot{\rho}=\mathcal{D}_{0}\rho+\mathcal{D}_{2}\rho+\mathcal{D}_{4}\rho$, where $\mathcal{D}_{0}\rho+\mathcal{D}_{2}\rho$ is  given in Eq.~\eqref{gamma0}, and $\mathcal{D}_{4}\rho$ is $4^{\text{th}}$-order in $\myhat{H}_{I}$. Crucially, the diagrammatic expansion  ensures the terms in $\mathcal{D}_{4}\rho$ are not separable---i.e. they correspond to genuine $4^{\text{th}}$-order processes, not products of $2^{\text{nd}}$-order terms.

\begin{figure}[htpb]
\begin{centering}
 \includegraphics[width=\columnwidth]{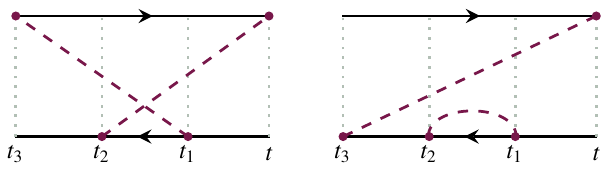}
 \par\end{centering}
  \caption{Example diagrams at $4^{\text{th}}$ order. The solid black lines are periods of free evolution of the system, interrupted by the action of the interaction Hamiltonian at times $t_{1},t_{2},t_{3}$ (purple dots). These  vertices are connected by purple dashed lines representing the cavity mode correlation functions. }
\label{Keldysh_3}
\end{figure}

Figure~\ref{Keldysh_3} shows two examples of $4^{\text{th}}$-order diagrams.
As described in Ref.~\cite{Muller2017:Deriving}, the solid horizontal lines indicate the branches of the Keldysh contour. The system undergoes free evolution along these branches, interrupted by the action of $H_I$ at times $t>t_1>t_2>t_3$ (purple dots). Points on the lower branch correspond to operators  to the left of the density matrix, while those on the upper branch are to the right.
Because the bath is quadratic, expectations of bath operators factorize into pairwise correlations (purple dashed lines).  The form of $H_I$ means each dashed line must connect opposite $\myhat{S}^\pm$ operators.  
Following these rules, after integrating over $t_1,t_2,t_3$ the diagrams in Fig.~\ref{Keldysh_3} correspond to:
\begin{multline}
4g^{4}\big[|\QA|^{2}\kappa^{-1}\myhat{S}^{-}\myhat{S}^{-}\rho\myhat{S}^{+}\myhat{S}^{+}+|\QB|^{2}\kappa^{-1}\myhat{S}^{+}\myhat{S}^{+}\rho\myhat{S}^{-}\myhat{S}^{-}\\+
\QA\QB^{*}Q_{\Delta}\myhat{S}^{+}\myhat{S}^{-}\rho\myhat{S}^{-}\myhat{S}^{+}+\QA^{*}\QB Q_{\Delta}^{*}\myhat{S}^{-}\myhat{S}^{+}\rho\myhat{S}^{+}\myhat{S}^{-}\\-
\QB^{2}Q_{\Sigma}\myhat{S}^{+}\myhat{S}^{-}\myhat{S}^{+}\rho\myhat{S}^{-}-\QA^{2}Q_{\Sigma}\myhat{S}^{-}\myhat{S}^{+}\myhat{S}^{-}\rho\myhat{S}^{+}\\-
\QA^{3}\myhat{S}^{+}\myhat{S}^{-}\myhat{S}^{-}\rho\myhat{S}^{+}-\QB^{3}\myhat{S}^{-}\myhat{S}^{+}\myhat{S}^{+}\rho\myhat{S}^{-}\big],
\label{fourthorder}
\end{multline}
with  $Q_{\Delta}=\left[\kappa+i\left(\omega_{A}-\omega_{B}-2\omega_{0}\right)\right]^{-1}$, and $Q_{\Sigma}=\left[\kappa+i\left(\omega_{A}+\omega_{B}\right)\right]^{-1}$.
Overall,  at $4^{\text{th}}$ order, there are $32$ diagrams. Considering the patterns of $S^\pm$ operators, each diagram contributes $4$ terms.  These  are written in full in the supplemental material~\cite{SM}. 

The resulting equation is not of Lindblad form~\cite{Lindblad1976b} and so does not necessarily preserve positivity---this is shown in~\cite{SM}, where we diagonalize the Lindblad--Kossakowski matrix.  As discussed in many other contexts, such non-positive equations can nonetheless predict correct behavior~\cite{Jeske2014:BlochRedfield,Eastham2016,Cammack2018,Dodin2018,Damanet2019:AtomOnly,Hartmann2020:accuracy}.  In $2^{\text{nd}}$-order theories, a Lindblad form equation is known to arise from secularizating the Redfield equation~\cite{Duemcke1979:Proper}---i.e., deleting terms which are time dependent in the interaction picture.  Since our equation is $U(1)$ symmetric, it already has a secularized form.

\begin{figure}[thpb]
\begin{centering}
 \hspace*{-.3cm}\includegraphics[width=1.05\columnwidth]{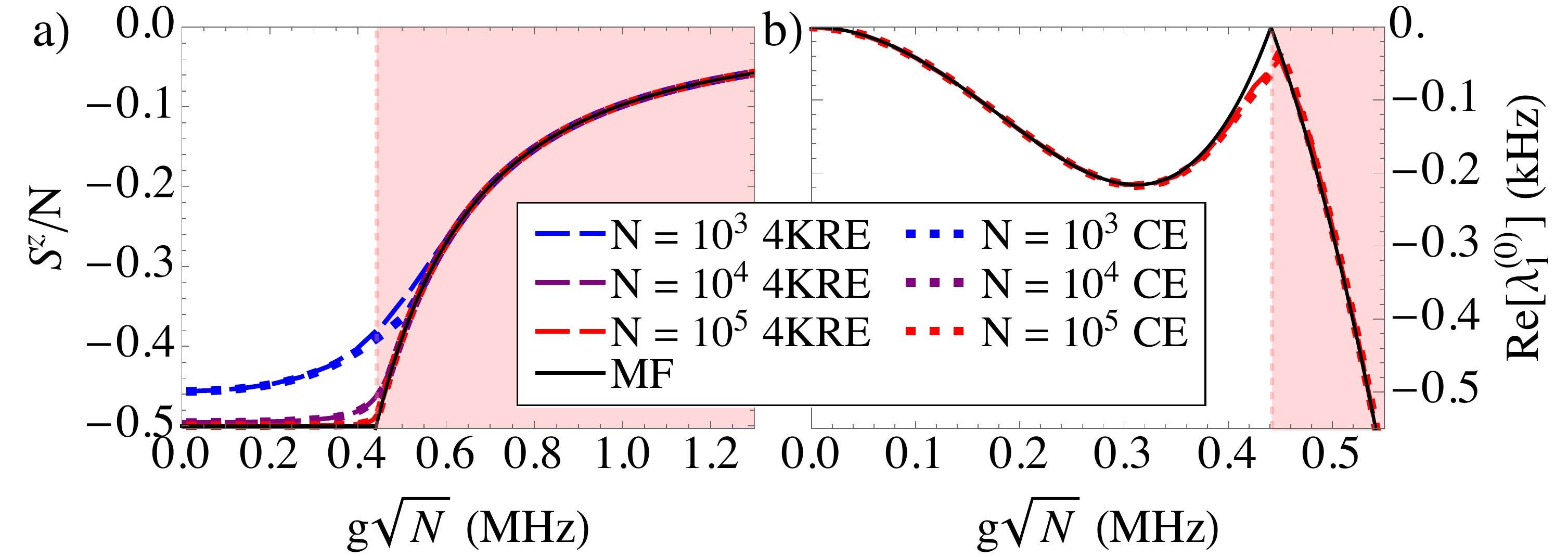}
  \par\end{centering}
  \caption{(a) Steady-state$\langle S^z \rangle$ evaluated from 4KRE (dashed lines) for three system sizes $N=10^3,10^4,10^5$. The results are compared to the cumulant equations (dotted lines), and mean-field theory of the atom-cavity equations (solid black line). (b) The first non-zero eigenvalue of: the $k=0$ sector of $\mathcal{L}$ for $N=10^5$ (dashed line), the linearized CE (dotted line) for $N=10^5$, and  the mean-field atom-cavity equations (black solid line). These eigenvalues describe relaxation towards the stationary state. The pink region indicates the superradiant phase, $g>g_{c}\sqrt{N}=0.44$ MHz for the set of parameters used: $\kappa=8.1$ MHz, $\omega_{0}=47$ kHz, $\omega_{A}=\omega_{B}=5$ MHz.}
\label{SS}
\end{figure}

While $U(1)$ symmetry still means the density matrix is diagonal in the $\myhat{S}^z$ basis, the presence of the $4^{\text{th}}$ order contribution gives a non-trivial dependence on $g$. In particular, as shown in Fig.~\ref{SS}(a), the steady state of this $4^{\text{th}}$-order Keldysh--Redfield equation (4KRE) shows a transition to a superradiant state. As $N$ increases, the results converge to the mean-field predictions of the full atom-cavity model, with
a discontinuity in $\langle S^z \rangle$ at $g=g_c$. 

\textit{Semiclassical equations}--- 
We now consider whether we can use the 4KRE to derive semiclassical equations which capture the dissipative phase transition. A naive approach to this is to assume  that terms $\langle \myhat{S}^\alpha \myhat{S}^\beta \myhat{S}^\gamma\rangle$ can be factorized. Applying such an approach directly to 4KRE fails, in the sense that after factorization, this yields equations for which $\langle \myhat{S}^{x,y}\rangle=0, \langle \myhat{S}^z \rangle = -N/2$ is always a stable solution, in contrast to the clear instability seen in the full quantum solution.  The structure of the full quantum solution (i.e., a density matrix diagonal in the $\myhat{S}^z$ basis) suggests the origin of this failure:  while $U(1)$ symmetry ensures $\langle \myhat{S}^{x,y} \rangle$ is zero for the full solution, products such as $\langle (\myhat{S}^{x})^2 \rangle$ can be non-zero.  The diagonal structure of the steady-state density matrix suggests a better way to approach a semiclassical limit is by considering the Gaussian form of the probabilities $P_M$ (see~\cite{SM}), and characterising $P_M$ by its first two moments.  This yields a form of cumulant equation (CE), as used elsewhere~\cite{gardiner2009stochastic,Kirton2017:Suppressing}.  Considering coupled equations for $\langle \myhat{S}^{z} \myhat{S}^{z} \rangle$, $\langle \myhat{S}^{z}  \rangle$ (given in~\cite{SM}), one finds results (dotted lines in Fig.~\ref{SS}(a)) that match  4KRE well for a range of $N$.  At $N\to \infty$,  $P_M$ is sharply peaked so $\langle S^zS^z\rangle\rightarrow\langle S^z\rangle^2$,  giving a single equation for $\langle \myhat{S}^z \rangle$:
\begin{equation}
\begin{split} 
\partial_t \langle {S}^{z} \rangle=\left[-2g^{2} \eta+2g^{4}\zeta\left\langle S^{z}\right\rangle\right]\left[S^2-\left\langle S^{z}\right\rangle^{2}\right],
\end{split}
\label{MFlimit}
\end{equation}
where $\eta=2\Re[\QA - \QB]$ and
$\zeta=8 \Re[(\QA+\QB)^{2}\QS]
- 16 \kappa^{-1} [ \Re(\QA)^2 + \Re(\QB)^2]$.

\textit{Liouvillian spectrum}--- 
To explore whether the atom-only equations we have derived correctly capture the slow dynamics, we next consider the Liouvillian spectrum. Due to the $U(1)$ symmetry, the quantum dynamics decouples into sectors labeled by index $k$, 
$\rho=
\sum_{k=-S}^S \sum_{M=-S-\min(k,0)}^{S-\min(k,0)}
R^{(k)}_M \ket{M}\bra{M+k}$.  The equations for each $k$ are independent:
$ \dot{R}^{(k)}_{M}={L}^{(k)}_{M,M^{\prime}}R^{(k)}_{M^{\prime}}$.
The matrices ${L}^{(k)}$ in each sector can be numerically diagonalized to obtain  eigenvalues $\lambda^{(k)}_{i}$. The real part of these eigenvalues describe the relaxation rate toward the steady state.  For our $4^{\text{th}}$-order approach, the matrices ${L}^{(k)}$ take a simple structure---only terms with $M^\prime = M, M\pm1, M\pm2$ are non-zero, yielding a pentadiagonal matrix~\cite{SM}.  Together, the separation into sectors and this banded structure mean we can numerically find the eigenspectrum for relatively large $N$.

The $k=0$ sector describes the dynamics of the populations (in the $\myhat{S}^z$ basis). As such, the eigenvalues in this sector provide information about the  evolution of $\langle \myhat{S}^z\rangle$, and in particular, the damping rate towards steady state. The first non-zero eigenvalue~\footnote{We label eigenvalues as: $\lambda^{(k)}_{i}$, with $i=0,1,2,\ldots$. Hence, $\lambda^{(0)}_{0}=0$ is the zero-mode associated to the steady state and $\lambda^{(0)}_{1}$ is the first non-zero eigenvalue in the $k=0$ sector.} in the $k=0$ sector, $\lambda^{(0)}_{1}$, is shown in Fig.~\ref{SS}(b). 
We can also compare this eigenvalue to the decay rate found by linearizing the semiclassical Eq.~\eqref{MFlimit}:
\begin{equation}
\lambda_{MF}=4g^{2}\eta \langle S^{z}\rangle_{ss} +
2g^{4}\zeta \left(S^2 -
3 \langle S^{z}\rangle_{ss}^{2}\right),
\label{MFeigenvalue}
\end{equation}
shown as the black solid line in Fig.~\ref{SS}(b).

While the first non-zero eigenvalue for $k=0$ tells us about relaxation to the steady state, it does not describe the slowest dynamics of this $U(1)$ symmetric system---i.e., the smallest non-zero eigenvalue, also known as the Liouvillian gap.  
As explained in Refs.~\cite{Minganti2018:Spectral}, when an open system spontaneously breaks a symmetry, the Liouvillian gap should vanish as $N\to\infty$ throughout the symmetry-broken phase.  This occurs because spontaneous symmetry breaking  means more than one steady state is possible;  since any mixture of two steady-state density matrices is also a steady state, an extra zero mode must arise. When the spontaneously broken symmetry is continuous, this also relates to the Goldstone mode.

\begin{figure}[htpb]
\begin{centering}
 \hspace*{-.3cm}\includegraphics[width=1.04\columnwidth]{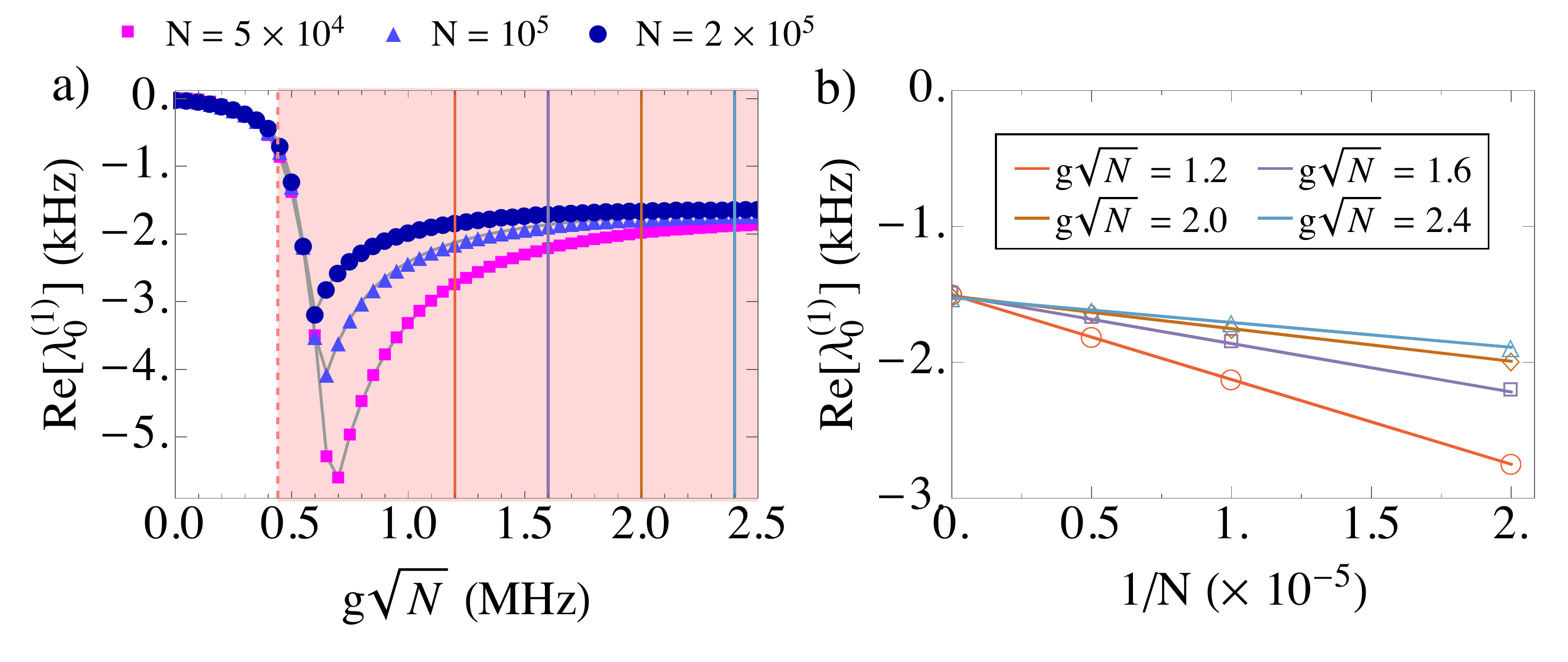}
 \par\end{centering}
  \caption{Liouvillian gap in the $|k|=1$ sector. a) Gap vs $g\sqrt{N}$.  The colored vertical lines indicate $g\sqrt{N}$ values above threshold used to evaluate the trend of the gap closure with the system size shown in b). This is computed for $N$ in the range $\left[5\times{10}^4-2\times 10^5\right]$. The intercept shows the extrapolated gap as $N\to\infty$.  All parameters as for Fig.~\ref{SS}}
\label{figure4}
\end{figure}

For our model, the gapless mode is associated with how the $U(1)$ symmetry is broken. As such, it must involve terms which are off-diagonal in the $S^z$ basis---specifically terms which correspond to the long-time evolution of $\langle \myhat S^+(t) \myhat S^-(0) \rangle$. This means we should consider the smallest eigenvalue (by real part) of the $|k|=1$ sector, $\lambda^{(1)}_{0}$.  This eigenvalue is shown in Fig.~\ref{figure4}(a), as a function of coupling $g$. One may see that the gap reduces with increasing $N$. However, a detailed analysis of the Liouvillian gap as a function of system size, Fig.~\ref{figure4}(b), shows a non-zero gap remains at $N\rightarrow\infty$. Specifically, considering $g>g_c$, we see this eigenvalue matches well to $\lambda^{(1)}_0 = A+B/N$, with a finite intercept $A$. As noted above, the 4KRE is not of Lindblad form (i.e. is not completelely positive), and it is possible this may be associated with the non-vanishing Liouvillian gap.  However, the equation is Hermitian, trace-preserving, respects the $U(1)$ symmetry, and predicts the correct steady states.  
Moreover, other examples of non-positive Redfield equations do show the expected vanishing Liouvillian gap, as shown in~\cite{SM}.

\textit{Conclusion}---
We have shown that constructing an atom-only effective theory for a model with U(1) symmetry presents surprising challenges. We have shown why standard $2^{\text{nd}}$ order Redfield theory fails to describe the superradiant transition, and we have shown how this can be rectified by the $4^{\text{th}}$ order terms of a diagrammatic expansion. We see this correctly describes the steady state and relaxation to that state in the $k=0$ sector.  Moreover, we may see how a semiclassical approximation becomes valid as $N\to\infty$, through the emergence of an increasingly sharp distribution---such an approach may provide alternate ways to understand models where it is found that the semiclassical (mean-field) approximation fails~\cite{Buca2019:Dissipation,Chiacchio2019:Dissipation}.  When considering spin coherences (i.e. the $k=1$ sector), we surprisingly find that  the 4KRE does not predict a vanishing Liouvillian gap in the symmetry-broken state.  Moreover, we note that the 4KRE provides a Hermitian, trace-preserving, and secularized density matrix equation which is nevertheless not of Lindblad form.  A key task for future work is to extend the methods developed here to a full multimode (e.g. confocal cavity) and spatially extended system.    This would provide a powerful tool to theoretically explore non-equilibrium phase transitions in these driven-dissipative multimode systems.

\textit{Acknowledgments}---
We are grateful for helpful discussions with A.~Daley, F.~Damanet. R.P. wishes to thank G.~Baio for stimulating discussions.  We are grateful to B.~Lev for comments on an earlier version of the manuscript.
R.P. was supported by the EPSRC Scottish Doctoral Training Centre in Condensed Matter Physics (CM-CDT), Grant number: EP/L015110/1.

%

\onecolumngrid
\clearpage

\renewcommand{\theequation}{S\arabic{equation}}
\renewcommand{\thefigure}{S\arabic{figure}}
\setcounter{equation}{0}
\setcounter{figure}{0}
\setcounter{page}{1}

\newcommand{\ZZ}{\left\langle S^{z}S^{z} \right\rangle}
\newcommand{\Z}{\left\langle S^{z} \right\rangle}
\newcommand{\posp}{\,\,\quad}
\newcommand{\nesp}{\hspace{-1.5em}}

\section{Supplemental Material for: ``Atom-only theories for U(1) symmetric cavity-QED models''}
\twocolumngrid

\section{Fourth order Keldysh-Redfield theory}

\subsection{Derivation of density matrix equation of motion}

\begin{figure*}[bht]
\begin{centering}
  \includegraphics[width=\textwidth]{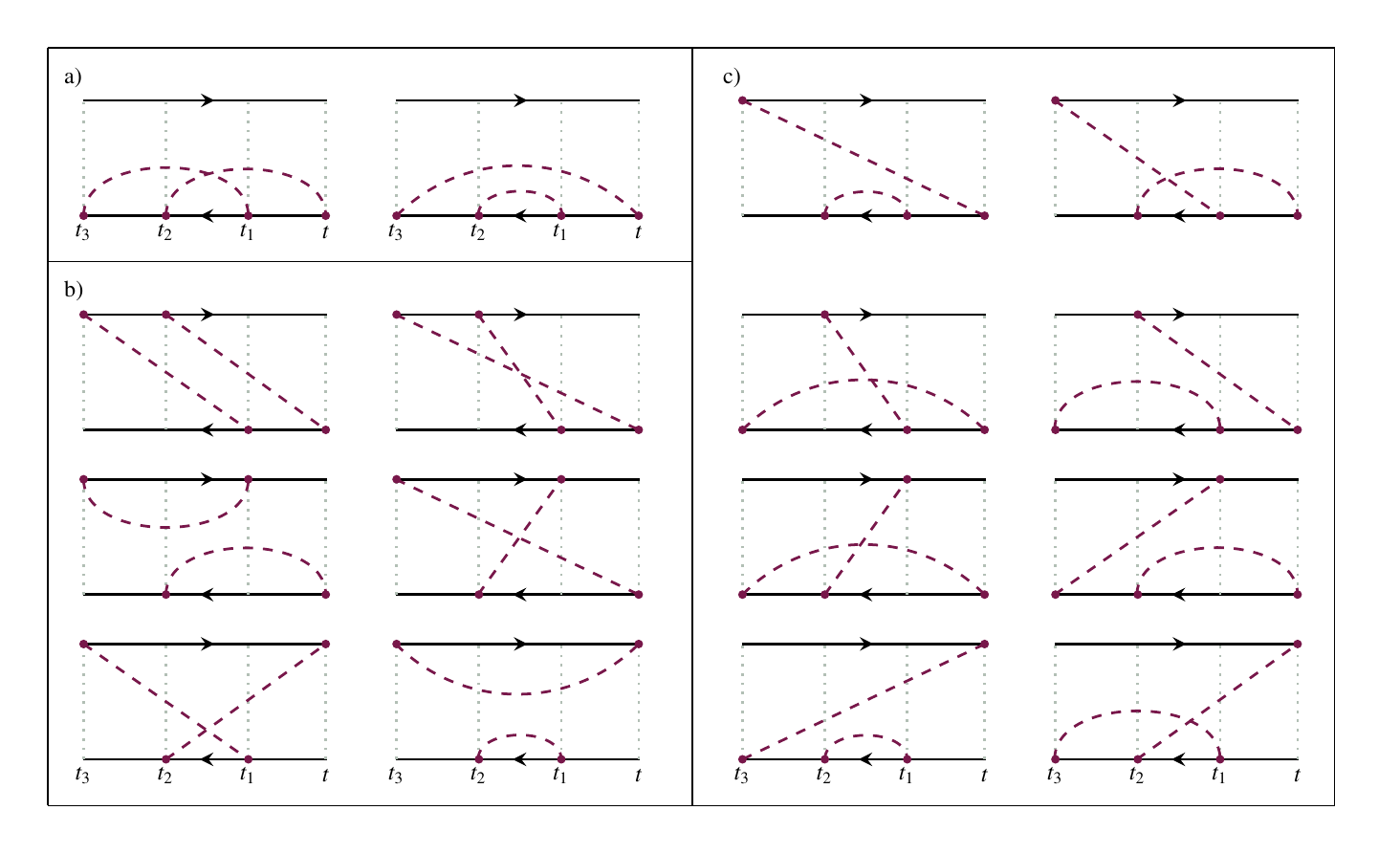}
  \par\end{centering}
  \caption{Keldysh diagrams at $4^{th}$ order.  Diagrams divide into three classes (a,b,c), according to the number of operators appearing to the left and right of the density matrix.  An additional 16 diagrams arise from inverting those drawn here.}
\label{diagrams}
\end{figure*}

This section provides details of the fourth order Keldysh-Redfield equation (4KRE), using Keldysh diagrammatic perturbation theory introduced by \citet{Muller2017:Deriving}. 

The quantum dynamics of the system density matrix, $\rho(t)$,  written in the interaction picture, is given by:
\begin{equation}
\dot{\rho}(t)=\int^{t}_{t_0}dt_1\rho(t_1)\Sigma(t_1,t)
\label{selfenergy}
\end{equation}
where $\Sigma(t_1,t)$ is a self-energy superoperator.  \citet{Muller2017:Deriving} provide a diagrammatic recipe to derive this self energy perturbatively in the system bath coupling (which here means $g$).  We use this approach up to fourth order to find $\Sigma(t_1,t)$, and then make a Markovian approximation $\rho(t_1)\simeq\rho(t)$ to provide a time-local equation of motion for $\rho$.

Because we consider evolution of a density matrix, the diagrammatic representation of the super-operators contributing to the self-energy consists of two horizontal solid lines representing  free evolution of the system to the left and right of the density matrix (bottom and top branches respectively).  An $n$th order diagram then contains $n$ vertices corresponding to the action of $\myhat{H}_{I}$ at the  times $t_1,...t_{n-1}$ and at $t$.  Dashed lines connecting these vertices represent pairwise correlations of the (Gaussian) bath, following Wick's theorem.  Crucially, only irreducible diagrams are drawn, describing the contributions of $n$th order terms which cannot be factorised into products of lower order terms.

The irreducible diagrams shown in Fig.~\ref{diagrams} are the fourth order contributions.  (The second order contributions are the standard Redfield theory, and their contribution $\mathcal{D}_0\rho+\mathcal{D}_2\rho$ is given explicitly in the Letter). In total there are $32$ fourth order diagrams, $16$ shown in Fig.~\ref{diagrams}, and an additional $16$ diagrams obtained by swapping vertices between the lower and upper branches. The diagrams can be divided in three categories: (a) terms where all operators act to the left of $\rho$, (b) terms where operators appear in pairs to the left and the right, and (c) terms with three operators to the left and one to the right. As stated in the Letter, the form of our interaction Hamiltonian means that vertices which are connected must have opposite spin flip operators $S^\pm$.
\begin{widetext}
  Evaluating these diagrams, the $4$th order contribution to the density matrix equation takes the form:
\begin{align} 
\mathcal{D}_4{\rho}=4g^{4}\Big[&Q^{3}_{-}\big(2\myhat{S}^{-}\myhat{S}^{-}\rho\myhat{S}^{+}\myhat{S}^{+}+2\myhat{S}^{+}\myhat{S}^{+}\myhat{S}^{-}\myhat{S}^{-}\rho-4\myhat{S}^{+}\myhat{S}^{-}\myhat{S}^{-}\rho\myhat{S}^{+}\big)+\nonumber\\
&Q^{3}_{+}\big(2\myhat{S}^{+}\myhat{S}^{+}\rho\myhat{S}^{-}\myhat{S}^{-}+2\myhat{S}^{-}\myhat{S}^{-}\myhat{S}^{+}\myhat{S}^{+}\rho-4\myhat{S}^{-}\myhat{S}^{+}\myhat{S}^{+}\rho\myhat{S}^{-}\big)+\nonumber\\
&Q_{-}Q_{+}Q_{\Sigma}\big(\myhat{S}^{+}\myhat{S}^{-}\rho\myhat{S}^{+}\myhat{S}^{-}+\myhat{S}^{-}\myhat{S}^{+}\rho\myhat{S}^{-}\myhat{S}^{+}+\myhat{S}^{+}\myhat{S}^{-}\myhat{S}^{-}\myhat{S}^{+}\rho+\myhat{S}^{-}\myhat{S}^{+}\myhat{S}^{+}\myhat{S}^{-}\rho+\nonumber\\
&\qquad\qquad-\myhat{S}^{+}\myhat{S}^{-}\myhat{S}^{+}\rho\myhat{S}^{-}-\myhat{S}^{-}\myhat{S}^{+}\myhat{S}^{-}\rho\myhat{S}^{+}-\myhat{S}^{-}\myhat{S}^{-}\myhat{S}^{+}\rho\myhat{S}^{+}-\myhat{S}^{+}\myhat{S}^{+}\myhat{S}^{-}\rho\myhat{S}^{-}\big)+\nonumber\\
&Q^{2}_{-}Q_{\Sigma}\big(\myhat{S}^{+}\myhat{S}^{-}\myhat{S}^{+}\myhat{S}^{-}\rho+\myhat{S}^{+}\myhat{S}^{-}\rho\myhat{S}^{-}\myhat{S}^{+}-\myhat{S}^{+}\myhat{S}^{+}\myhat{S}^{-}\rho\myhat{S}^{-}-\myhat{S}^{-}\myhat{S}^{+}\myhat{S}^{-}\rho\myhat{S}^{+}\big)+\nonumber\\
&Q^{2}_{+}Q_{\Sigma}\big(\myhat{S}^{-}\myhat{S}^{+}\myhat{S}^{-}\myhat{S}^{+}\rho+\myhat{S}^{-}\myhat{S}^{+}\rho\myhat{S}^{+}\myhat{S}^{-}-\myhat{S}^{-}\myhat{S}^{-}\myhat{S}^{+}\rho\myhat{S}^{+}-\myhat{S}^{+}\myhat{S}^{-}\myhat{S}^{+}\rho\myhat{S}^{-}\big)+\nonumber\\
&\frac{|Q_{-}|^2}{\kappa}\big(\myhat{S}^{+}\myhat{S}^{-}\rho\myhat{S}^{+}\myhat{S}^{-}+\myhat{S}^{-}\myhat{S}^{-}\rho\myhat{S}^{+}\myhat{S}^{+}-\myhat{S}^{+}\myhat{S}^{-}\myhat{S}^{-}\rho\myhat{S}^{+}-\myhat{S}^{-}\myhat{S}^{+}\myhat{S}^{-}\rho\myhat{S}^{+}
\big)+\nonumber\\
&\frac{|Q_{+}|^2}{\kappa}\big(\myhat{S}^{-}\myhat{S}^{+}\rho\myhat{S}^{-}\myhat{S}^{+}+\myhat{S}^{+}\myhat{S}^{+}\rho\myhat{S}^{-}\myhat{S}^{-}-\myhat{S}^{-}\myhat{S}^{+}\myhat{S}^{+}\rho\myhat{S}^{-}-\myhat{S}^{+}\myhat{S}^{-}\myhat{S}^{+}\rho\myhat{S}^{-}
\big)+\nonumber\\
&Q_{-}Q^{\ast}_{+}Q_{\Delta}\big(4\myhat{S}^{+}\myhat{S}^{-}\rho\myhat{S}^{-}\myhat{S}^{+}-2\myhat{S}^{+}\myhat{S}^{+}\myhat{S}^{-}\rho\myhat{S}^{-}-2\myhat{S}^{-}\rho\myhat{S}^{-}\myhat{S}^{+}\myhat{S}^{+}\big)+\nonumber\\
&\kappa^{-1}{Q^{2}_{-}}\big(\myhat{S}^{-}\myhat{S}^{-}\rho\myhat{S}^{+}\myhat{S}^{+}+\myhat{S}^{+}\myhat{S}^{-}\rho\myhat{S}^{+}\myhat{S}^{-}-\myhat{S}^{-}\rho\myhat{S}^{+}\myhat{S}^{-}\myhat{S}^{+}-\myhat{S}^{+}\myhat{S}^{-}\myhat{S}^{-}\rho\myhat{S}^{+}\big)+\nonumber\\
&\kappa^{-1}{Q^{2}_{+}}\big(\myhat{S}^{+}\myhat{S}^{+}\rho\myhat{S}^{-}\myhat{S}^{-}+\myhat{S}^{-}\myhat{S}^{+}\rho\myhat{S}^{-}\myhat{S}^{+}-\myhat{S}^{+}\rho\myhat{S}^{-}\myhat{S}^{+}\myhat{S}^{-}-\myhat{S}^{-}\myhat{S}^{+}\myhat{S}^{+}\rho\myhat{S}^{-}\big)+\nonumber\\
&Q^{2}_{-}Q_{\Delta}\big(2\myhat{S}^{+}\myhat{S}^{-}\rho\myhat{S}^{-}\myhat{S}^{+}-\myhat{S}^{-}\rho\myhat{S}^{-}\myhat{S}^{+}\myhat{S}^{+}-\myhat{S}^{+}\myhat{S}^{+}\myhat{S}^{-}\rho\myhat{S}^{-}\big)+\nonumber\\
&Q^{2}_{+}Q^{\ast}_{\Delta}\big(2\myhat{S}^{-}\myhat{S}^{+}\rho\myhat{S}^{+}\myhat{S}^{-}-\myhat{S}^{+}\rho\myhat{S}^{+}\myhat{S}^{-}\myhat{S}^{-}-\myhat{S}^{-}\myhat{S}^{-}\myhat{S}^{+}\rho\myhat{S}^{+}\big)+\text{H.c}\Big]
\label{KRE4}
\end{align}
where we have used $Q_{\mp}=\left[\kappa+2i\left(\omega_{A(B)}\mp\omega_{0}\right)\right]^{-1}$, $Q_{\Sigma}=\left[\kappa+i\left(\omega_{A}+\omega_{B}\right)\right]^{-1}$, $Q_{\Delta}=\left[\kappa+i\left(\omega_{A}-\omega_{B}-2\omega_{0}\right)\right]^{-1}$. 
\end{widetext}

\subsection{Lindblad form of density matrix equation}

By construction, Eq.~\eqref{KRE4} satisfies  Hermiticity and trace preservation. In addition, one may see that it respects the U(1) symmetry as there are always equal numbers of raising and lowering operators.  In this section we explore whether this structure also preserves positivity, equivalent to requiring that the density matrix equation of motion takes Lindblad form~\cite{Lindblad1976b}.  We note however that, as discussed elsewhere~\cite{Jeske2014a,Eastham2016,Cammack2018,Dodin2018,Damanet2019:AtomOnly}, there are many situations where a Redfield equation which does not preserve positivity may yet give an accurate description of the reduced system dynamics. To check for the Lindblad form we follow the method in Ref.~\cite{Breuer2002} to extract the Lindblad--Kossakowski matrix. 
We describe here the approach we use to perform this numerically.

To construct the Lindblad--Kossakowski matrix we start by writing our equation of motion, $\dot{\rho}=\mathcal{M}\rho$, in matrix form:
\begin{equation}
\dot{\rho}_{mn}=\mathcal{M}_{nmpq}\rho_{pq}.
\label{mematrix}
\end{equation}
The next step is to rewrite Eq.~\eqref{mematrix} in terms of a complete set of linearly independent $N\times N$ matrices spanning the Hilbert space.
It will be convenient to make use of two sets of basis matrices. The first is the element basis $O_i$, with $i=0,...N^2 -1$. These matrices are defined by $[O_i]_{j,k}=\delta_{j,Q_{i,N}}\delta_{k,R_{i,N}}$ where $Q_{i,N}$ and $R_{i,N}$ are respectively the quotient and the remainder of $i$ modulo $N$. The second set is the normalized generalized Gell--Mann (gGM) basis, $\gamma_i$, with $i=0,...N^2 -1$. These matrices are defined by $Tr(\gamma_i\gamma_j)=\delta_{ij}$. We include the identity matrix in this set as $\gamma_0=\mathbb{1}_N/\sqrt{N}$.  The next $N$ matrices, $p=1\ldots N$ are diagonal matrices, $\gamma_p=\text{diag}(1,1, \ldots, -p, 0, 0, \ldots)/\sqrt{p(1+p)}$  (i.e. matrices with first $p$ diagonal elements being one, the next element being $-p$, and all other elements zero).
The remaining matrices are off-diagonal, of the form:
\begin{equation}
[\gamma_{IJ}^x]_{ij}=\frac{1}{\sqrt{2}}(\delta_{iI}\delta_{jJ}+\delta_{iJ}\delta_{jI}),
[\gamma_{IJ}^y]_{ij}=\frac{i}{\sqrt{2}}(\delta_{iI}\delta_{jJ}-\delta_{iJ}\delta_{jI}),
\label{gellmannoff}
\end{equation}
labeled by ordered pairs of integers $I>J$.  We use two bases as it is simplest to translate $\mathcal{M}$ to the element basis, but to find the Lindblad--Kossakowski matrix we need a basis with the identity matrix as one of the elements.

Using the above, we first write the density matrix equation in terms of the element basis as
\begin{equation}
\dot{\rho}=\sum_{i,j}L^{O}_{ij}O_i\rho O^{\dagger}_j.
\label{step1}
\end{equation}
Summing explicitly  over the quotient and remainders of $i,j$,
and defining the function $I(Q,R)=QN+R$, yields:
\begin{align}
\dot{\rho}_{nm}&=\langle n|\sum_{QR,Q^{\prime}R^{\prime}}L^{O}_{I(Q,R),I(Q^{\prime},R^{\prime})}O_{QR}\rho O_{R^{\prime}Q^{\prime}}|m\rangle
\nonumber\\
&=\sum_{R,R^{\prime}}L^{O}_{I(n,R),I(m,R^{\prime})}\rho_{RR^{\prime}},
\label{firstStep}
\end{align}
Therefore, by comparison with Eq.~\eqref{mematrix}, we find:
\begin{equation}
L^{O}_{I(n,p),I(m,q)}=\mathcal{M}_{nmpq}
\label{Lmatrix1}
\end{equation}
Now we have $L^{O}$, the next step is to make a basis transformation to the gGM basis, using $O_i=X_{ij}\gamma_j$ where $X_{ij}=Tr(O_i\gamma_j)$.  We thus have:
\begin{multline}
\dot{\rho}=\sum_{ij}\sum_{kl}L^{O}_{kl}X_{ki}\gamma_i\rho[X_{lj}\gamma_j]^{\dagger}
\equiv\sum_{ij}L^{\gamma}_{ij}\gamma_i\rho\gamma_j
\label{secondStep}
\end{multline}
where we have used $\gamma^{\dagger}_{i}=\gamma_{i}$.  The matrix
$\mathbf{L}^\gamma$ thus has the form:
\begin{math}
\mathbf{L}^{\gamma}=\mathbf{X}^T\mathbf{L}^{O}\mathbf{X}^{\ast}.
\end{math}

Because the gGM basis contains the identity matrix as element $0$, we may now use standard results~\cite{Breuer2002} to write:
\begin{equation}
\dot{\rho}=-i[H,\rho]+\sum^{N^{2}-1}_{i,j=1}L^{\gamma}_{ij}\left(\gamma_i\rho\gamma_j-\frac{1}{2}\left\{\rho,\gamma_j\gamma_i\right\}\right)
\label{mefinal},
\end{equation}
where we have defined:
\begin{equation}
H=\sum^{N^{2}-1}_{i=1}\frac{L^{\gamma}_{0i}-L^{\gamma}_{i0}}{2i\sqrt{N}}\gamma^i,
\label{HforLform}
\end{equation}
We thus see that the submatrix of
$\mathbf{L}^{\gamma}$  excluding row and column $0$ is the Lindblad--Kossakowski matrix.  Extracting this numerically from our density matrix equation of motion, we find that the 4KRE is not of Lindblad form since not all  eigenvalues are positive. We also note that independent  of $N$, we find only $8$ non-zero eigenvalues.

To conclude this section, we note that for second-order Redfield equations, one way to put the  equation into Lindblad form is secularization~\cite{Duemcke1979:Proper}: eliminating from the Redfield equation those terms which are time dependent in the interaction picture.  For our equation, the $U(1)$ symmetry (and thus matching numbers of raising and lowering operators) automatically means that the equation is time-independent in the interaction picture. As such, it is notable that the 4KRE yields an equation for which secularization does not ensure positivity.

\subsection{Block structure of density matrix equation}

In this section, we analyze the block structure of Eq.~\eqref{KRE4}. As noted in the Letter the U(1) symmetry means the equation of motion breaks up into separate blocks.  By writing
\begin{equation}
 \rho=
\sum_{k=-S}^S \sum_{M=-S-\min(k,0)}^{S-\min(k,0)}
R^{(k)}_M \ket{M}\bra{M+k}   
\end{equation}
we find that each $k$ block evolves independently:
\begin{equation}
\dot{R}^{(k)}_{M}={L}^{(k)}_{M,M^{\prime}}R^{(k)}_{M^{\prime}}
\label{RKM}.
\end{equation}
We may thus also label eigenvalues and eigenvectors by blocks,
$L^{(k)}_{M,M^\prime} R^{(k)}_{n,M^\prime} = \lambda^{(k)}_n R^{(k)}_{n,M}$.

The matrix ${L}^{(k)}_{M,M^{\prime}}$takes a pentadiagonal form, which we may write as follows: 
\begin{multline}
{L}^{(k)}_{M,M^{\prime}}
=A^{(k)}_M \delta_{M,M^\prime}
+B^{(k)}_M \delta_{M,M^\prime-1}
+C^{(k)}_M \delta_{M,M^\prime-2}\\
+D^{(k)}_M \delta_{M,M^\prime+1}
+E^{(k)}_M \delta_{M,M^\prime+2}.
\label{L}
\end{multline}
Equivalently, in explicit matrix form, this means:
\begin{equation}
\mathbf{L}^{(k)}=\left(
\begin{array}{ccccc}
A^{(k)}_{-S} & B^{(k)}_{-S} & C^{(k)}_{-S} & 0 & 0\\

D^{(k)}_{-S+1} & \ddots & \ddots & \ddots & 0\\

E^{(k)}_{-S+2} & \ddots & \ddots & \ddots & C^{(k)}_{S-2-k}\\

0 & \ddots & \ddots & \ddots & B^{(k)}_{S-1-k}\\

0 & 0 & E^{(k)}_{S-k} & D^{(k)}_{S-k} & A^{(k)}_{S-k}
\end{array}.
\right)
\end{equation}

The $k=0$ sector addresses the behavior of the symmetry-preserving variables as $\langle S^z\rangle$ and $\langle S^zS^z\rangle$, which may have a non-zero steady state value.  In contrast $k\neq 0$ describes coherences, which (for finite $N$) must vanish at long times due to the $U(1)$ symmetry.
The spectrum of the $k=0$ block thus provides details about the relaxation dynamics to the stationary states $\langle S^z\rangle_{SS}$ and $\langle S^zS^z\rangle_{SS}$.

\begin{widetext}
The five functions $A^{(k)}_M \ldots E^{(k)}_M$ can be written straightforwardly in terms of the matrix elements of spin raising and lowering operators,
$f^{M}\equiv\sqrt{\left(S-M\right)\left(S+M+1\right)}$.  We thus find:

\begin{align*}
A^{(k)}_M
=&i\omega_0 k-2g^2\left[\QA(f^{M-1})^2 + \QA^{\ast}(f^{M+k-1})^2 + \QB(f^{M})^2 +\QB^{\ast}(f^{M+k})^2\right] \\
& +4g^4\Big[2(\QA)^3(f^{M-1})^2(f^{M-2})^2 + 2(\QA^{\ast})^3(f^{M+k-1})^2(f^{M+k-2})^2 + \\
& 2(\QB)^3(f^{M})^2(f^{M+1})^2 + 2(\QB^{\ast})^3(f^{M+k})^2(f^{M+k+1})^2 + \\
& \QA\QB\QS \left((f^{M-1})^2(f^{M+k-1})^2 + (f^{M})^2(f^{M+k})^2 + 2(f^{M})^2(f^{M-1})^2\right)  +  \\
& \QA^{\ast}\QB^{\ast}\QS^{\ast} \left((f^{M-1})^2(f^{M+k-1})^2 + (f^{M})^2(f^{M+k})^2 + 2(f^{M+k})^2(f^{M+k-1})^2\right)  +  \\
& (\QA)^2\QS \left((f^{M-1})^4 + (f^{M-1})^2(f^{M+k})^2 \right) + (\QA^{\ast})^2\QS^{\ast} \left((f^{M+k-1})^4 + (f^{M})^2(f^{M+k-1})^2 \right)  + \\
&  (\QB)^2\QS \left((f^{M})^4 + (f^{M})^2(f^{M+k-1})^2 \right) + (\QB^{\ast})^2\QS^{\ast} \left((f^{M+k})^4 + (f^{M+k})^2(f^{M-1})^2 \right) + \\
& 2\frac{|\QA|^2}{\kappa}(f^{M-1})^2(f^{M+k-1})^2 +2\frac{|\QB|^2}{\kappa}(f^{M})^2(f^{M+k})^2 +\\
& 4\QA\QB^{\ast}\QD (f^{M+k})^2(f^{M-1})^2 +  
4\QA^{\ast}\QB\QD^{\ast} (f^{M})^2(f^{M+k-1})^2 + \\
& \left(\frac{(\QA)^2+({\QA}^{\ast})^2}{\kappa} \right) (f^{M-1})^2(f^{M+k-1})^2  + 
\left(\frac{(\QB)^2+({\QB}^{\ast})^2}{\kappa}  \right) (f^{M})^2(f^{M+k})^2 +\\
& 2(\QA)^2\QD(f^{M+k})^2(f^{M-1})^2 + 2(\QA^{\ast})^2\QD^{\ast}(f^{M})^2(f^{M+k-1})^2 + 
\\& 2(\QB)^2\QD^{\ast}(f^{M})^2(f^{M+k-1})^2 + 2(\QB^{\ast})^2\QD(f^{M+k})^2(f^{M-1})^2  \Big],
\\[1ex]
B^{(k)}_M=&2g^2 \left(\QA + \QA^{\ast} \right)f^{M}f^{M+k}  
 -4g^4 \Big[4(\QA)^3f^{M}f^{M+k}(f^{M-1})^2 + 4(\QA^{\ast})^3f^{M}f^{M+k}(f^{M+k-1})^2 +\\
&\QA\QB\QS \left((f^{M})^3f^{M+k} + f^{M}f^{M+k}(f^{M+1})^2 \right) +  \QA^{\ast}\QB^{\ast}\QS^{\ast} \left(f^{M}(f^{M+k})^3 + f^{M}f^{M+k}(f^{M+k+1})^2 \right) +\\
& (\QA)^2\QS (f^{M})^3f^{M+k}  + 
(\QA^{\ast})^2\QS^{\ast} f^{M}(f^{M+k})^3 + 
(\QB)^2\QS f^{M}f^{M+k}(f^{M+1})^2 + 
(\QB^{\ast})^2\QS^{\ast} f^{M}f^{M+k}(f^{M+k+1})^2 + \\
& \frac{|\QA|^2}{\kappa}\left(f^{M}f^{M+k}(f^{M-1})^2 + f^{M}f^{M+k}(f^{M+k-1})^2 + (f^{M})^3f^{M+k} + f^{M}(f^{M+k})^3 \right) +  \\
& 2\QA\QB^{\ast}\QD f^{M}f^{M+k}(f^{M+k+1})^2 +2\QA^{\ast}\QB\QD^{\ast} f^{M}f^{M+k}(f^{M+1})^2 +  \\
& \frac{(\QA)^2}{\kappa}\left(f^{M}(f^{M+k})^3 + f^{M}f^{M+k}(f^{M-1})^2 \right) -\frac{({\QA}^{\ast})^2}{\kappa}\left((f^{M})^3f^{M+k} + f^{M}f^{M+k}(f^{M+k-1})^2 \right) +  \\
& (\QA)^2\QD f^{M}f^{M+k}(f^{M+k+1})^2 + 
(\QA^{\ast})^2\QD^{\ast} f^{M}f^{M+k}(f^{M+1})^2 + \\
& (\QB)^2\QD^{\ast}f^{M}f^{M+k}(f^{M+1})^2 + 
 (\QB^{\ast})^2\QD f^{M}f^{M+k}(f^{M+k+1})^2  \Big], 
\\[1ex]
C^{(k)}_M=&4g^4 \left[2(\QA)^3+2(\QA^{\ast})^3 +\frac{(\QA+\QA^{\ast})^2}{\kappa}\right]f^{M}f^{M+1}f^{M+k}f^{M+k+1},
\\[1ex]
D^{(k)}_M=& 2g^2 \left(\QB + \QB^{\ast} \right)f^{M-1}f^{M+k-1} -
4g^4 \Big[4(\QB)^3(f^{M})^2f^{M-1}f^{M+k-1} + 4(\QB^{\ast})^3(f^{M+k})^2f^{M-1}f^{M+k-1} +\\
& \QA\QB\QS \left((f^{M-1})^3f^{M+k-1} + f^{M-1}(f^{M-2})^2f^{M+k-1} \right) + 
\QA^{\ast}\QB^{\ast}\QS^{\ast} \left(f^{M-1}(f^{M+k-1})^3 + f^{M-1}f^{M+k-1}(f^{M+k-2})^2 \right) +\\
& (\QA)^2\QS f^{M-1}f^{M+k-1}(f^{M-2})^2 + 
(\QA^{\ast})^2\QS^{\ast} f^{M-1}f^{M+k-1}(f^{M+k-2})^2 + \\
&
(\QB)^2\QS (f^{M-1})^3f^{M+k-1} + (\QB^{\ast})^2\QS^{\ast} f^{M-1}(f^{M+k-1})^3 + \\
& \frac{|\QB|^2}{\kappa}\left((f^{M})^2f^{M-1}f^{M+k-1} + (f^{M+k})^2f^{M-1}f^{M+k-1} + (f^{M-1})^3f^{M+k-1} + f^{M-1}(f^{M+k-1})^3 \right) +  \\
& 2\QA\QB^{\ast}\QD f^{M-1}f^{M+k-1}(f^{M-2})^2 + 2\QA^{\ast}\QB\QD^{\ast} f^{M-1}f^{M+k-1}(f^{M+k-2})^2 +  \\
& \frac{(\QB)^2}{\kappa}\left(f^{M-1}(f^{M+k-1})^3 + (f^{M})^2f^{M-1}f^{M+k-1} \right) + 
 \frac{({\QB}^{\ast})^2}{\kappa}\left((f^{M-1})^3f^{M+k-1} + (f^{M+k})^2f^{M-1}f^{M+k-1} \right) +  \\
& (\QA)^2\QD f^{M-1}f^{M+k-1}(f^{M-2})^2 + (\QA^{\ast})^2\QD^{\ast} f^{M-1}f^{M+k-1}(f^{M+k-2})^2  + \\
& (\QB)^2\QD^{\ast}f^{M-1}f^{M+k-1}(f^{M+k-2})^2  + (\QB^{\ast})^2\QD f^{M-1}f^{M+k-1}(f^{M-2})^2  \Big],
\\[1ex]
E^{(k)}_M=&4g^4 \left[2(\QB)^3+2(\QB^{\ast})^3 +\frac{(\QB+\QB^{\ast})^2}{\kappa}\right]f^{M-1}f^{M-2}f^{M+k-1}f^{M+k-2}.
\end{align*}

\end{widetext}

\section{Large $N$ properties of 4KRE results}

\subsection{Convergence of $k=0$ results at large $N$}

\begin{figure}[htpb]
 \begin{centering}
   \includegraphics[width=\columnwidth]{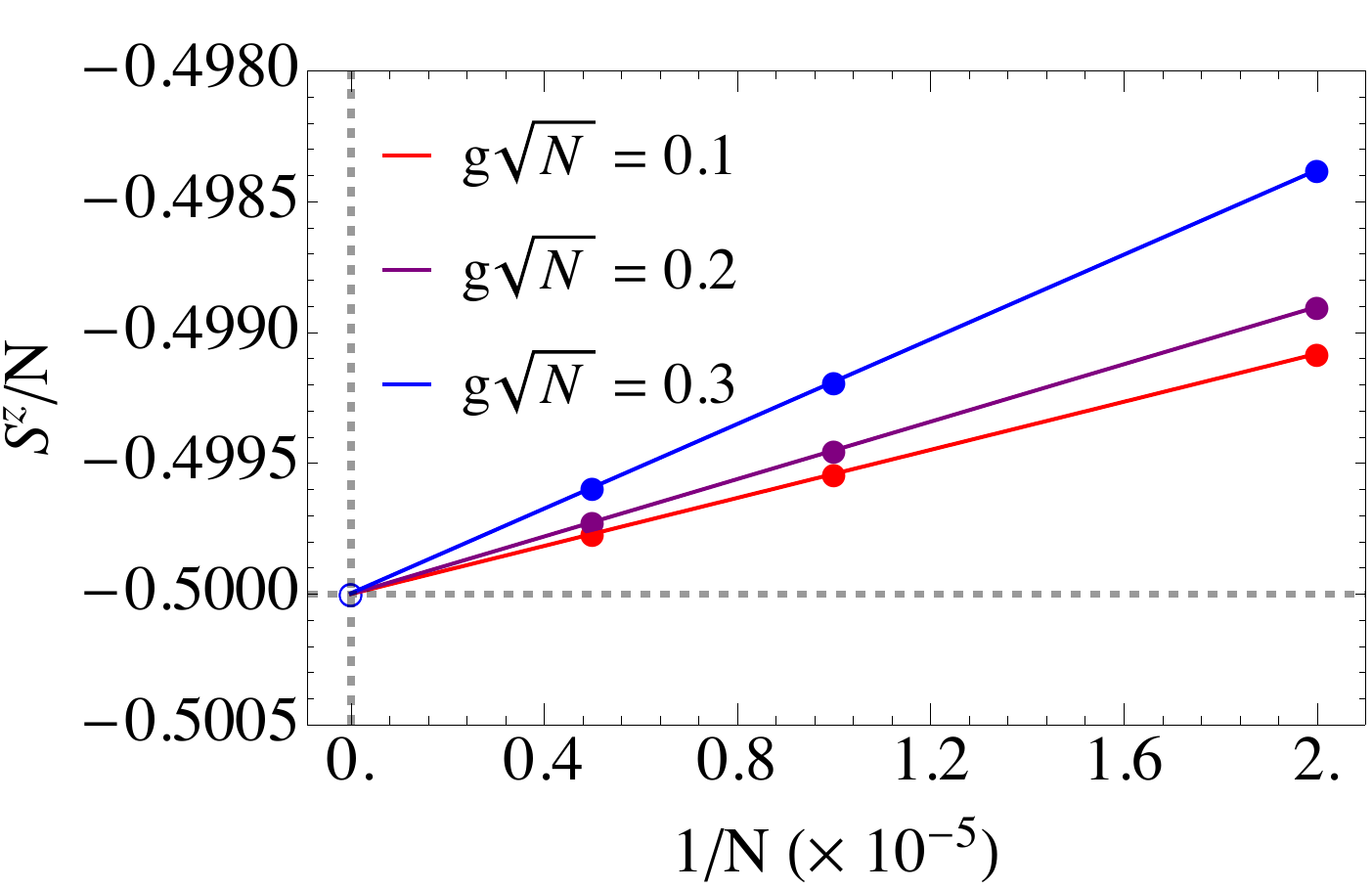}
   \par\end{centering}
   \caption{Scaling of the steady state value $\langle S^z \rangle$ with system size at three fixed values of coupling $g\sqrt{N}$ below threshold ($g_c\sqrt{N}=0.44$ MHz). The parameters used are $\kappa=8.1$ MHz, $\omega_{0}=47$ kHz, $\omega=5$ MHz.}
 \label{SteadyStateConvergence}
 \end{figure}

In the Letter we note that the Liouvillian gap, evaluated from the $k=\pm 1$ sector, does not vanish as expected at large $N$.  To verify there is no similar issue in the $k=0$ sector, Fig.~\ref{SteadyStateConvergence} shows how, for $g<g_c$ the value of $S^z/N$  converges to $-1/2$ as $N\to\infty$.

\subsection{Probability distribution at large $N$}

 \begin{figure*}[htb]
 \begin{centering}
   \includegraphics[width=\textwidth]{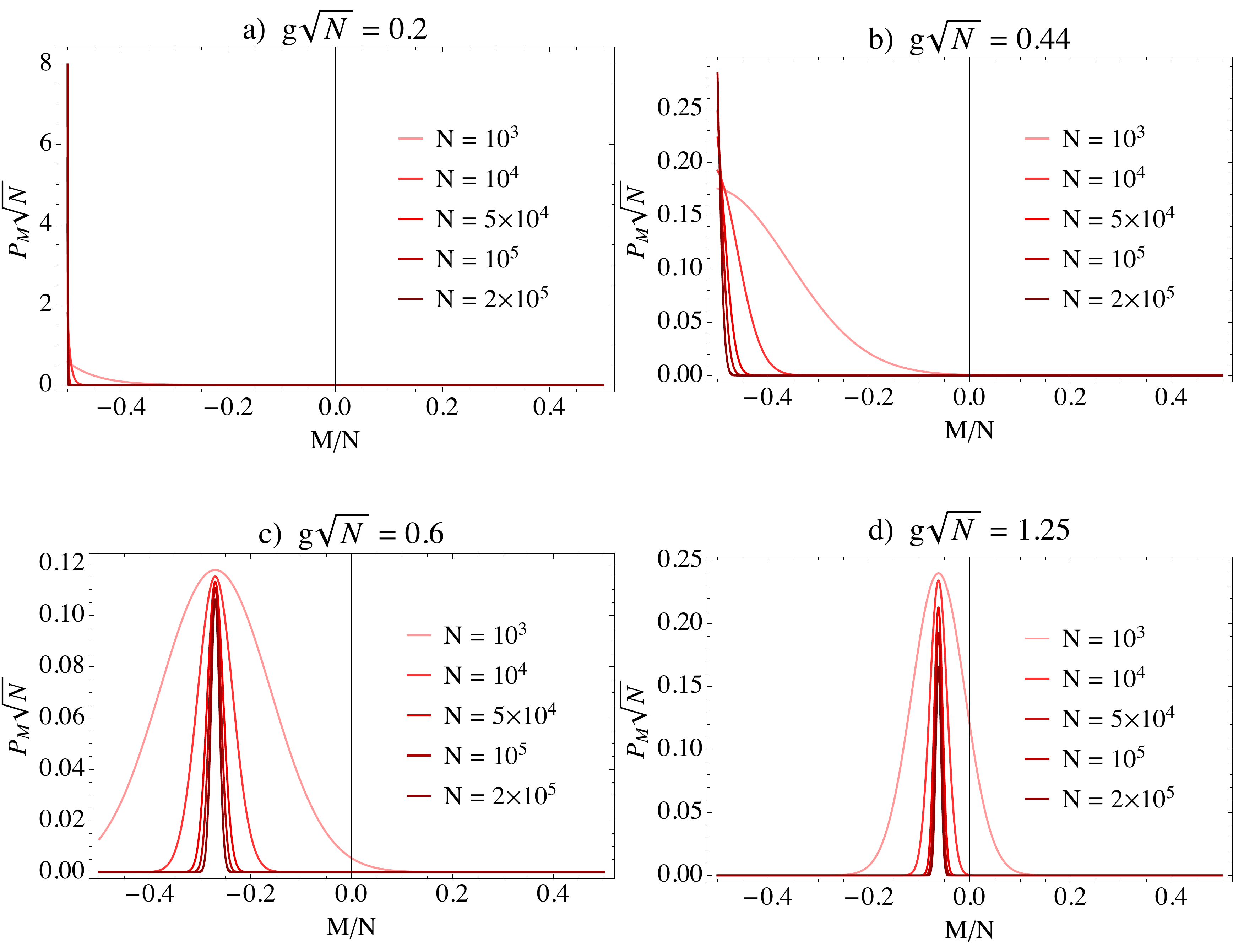}
   \par\end{centering}
   \caption{Profiles of the steady state probability distribution in function of the rescaled magnetic moment at fixed coupling strengths $g\sqrt{N}$. Panel a) is below threshold, $g<g_c$, panel b) at threshold, $g=g_c$ and panels c)-d) above threshold. The parameters used are $\kappa=8.1$ MHz, $\omega_{0}=47$ kHz, $\omega=5$ MHz.}
 \label{FigPD}
 \end{figure*}

In the Letter we discussed how the form of the probability $P_M$ suggested the approach of writing coupled equations for the first two moments of the probability distribution.  Here we present the results that support this statement.  
The probability distribution in the steady state is given by $P_M = R^{(0)}_{0,M}$ where $R^{(0)}_{0,M}$ is the eigenvector of Eq.~\eqref{L}, corresponding to eigenvalue $\lambda^{(0)}_{0}=0$.
Figure~\ref{FigPD} shows the probability distribution for four different values of $g\sqrt{N}$, and a range of values of $N$. Below threshold (panel a), for $g<g_c$, the distribution is peaked at $M=-N/2$, with a finite width that vanishes as $N$ increases, consistent with Fig.~\ref{SteadyStateConvergence}. In contrast, for $g>g_c$ the peak of the distribution moves away from $M=-N/2$,  and asymptotically approaches a peak at $M=0$ at large $g\sqrt{N}$ (panels c,d). 

One clearly sees that for $g>g_c$ the probability distribution has a quasi-Gaussian form, with a width that shrinks as $N$ increases.  For $g<g_c$, the probability distribution is one-sided, but can still be fit effectively by a Gaussian (peaked at $S^z<-N/2$).  These observations explain  why the cumulant equation adequately describes the steady state at finite $N$, and how  mean-field theory becomes valid in the thermodynamic limit.

\subsection{Cumulant equations of motion}

As we have seen, the steady-state probability distribution is approximately Gaussian, and so it should be possible to accurately model using its first two moments. We thus derive  equations of motion  for $\langle S^z\rangle$ and $\langle S^zS^z\rangle$. 
These equations depend on third order correlators of spin operators (from the second order terms in the density matrix equation) and fourth order correlators (from the fourth order terms).  Such terms can be decoupled to second order terms using the cumulant expansion~\cite{gardiner2009stochastic}.  

Because the cumulant expansion here is based on the approximate Gaussian distribution of $P_M$, such an expansion applies exclusively to decoupling products of $S^z$ operators.  As such, before decoupling the equations of motion, we first rewrite all expressions in terms of $S^z$ operators by using the identity 
$S^{+}S^{-}=S\left(S+1\right)-S^{z}S^{z}+S^{z}$.  The $U(1)$ symmetry of our model ensures that this will always be possible. We also note that we must keep track of non-commuting operators $S^\pm, S^z$ in this process.  Once all terms are written in terms of only $S^z$,  these operators commute, and so can be treated analogously to classical cumulant expansions.

\newcommand{\SpSm}{\left\langle S^{+}S^{-} \right\rangle}
\newcommand{\SmSp}{\left\langle S^{-}S^{+} \right\rangle}
\newcommand{\pzm}{\left\langle S^{+}S^{z}S^{-} \right\rangle}
\newcommand{\zpm}{\left\langle S^{z}S^{+}S^{-} \right\rangle}
\newcommand{\pmz}{\left\langle S^{+}S^{-}S^{z} \right\rangle}
\newcommand{\cpzm}{\left\langle \left[S^{+},S^{z}\right]S^{-} \right\rangle}
\newcommand{\czpm}{\left\langle \left[S^{z},S^{+}\right]S^{-} \right\rangle}
\newcommand{\pzzm}{\left\langle S^{+}{S^{z}}^2S^{-} \right\rangle}
\newcommand{\pmzz}{\left\langle S^{+}S^{-}{S^{z}}^2 \right\rangle}
\newcommand{\pczzm}{\left\langle S^{+}\left[{S^{z}}^2,S^{-}\right] \right\rangle}
\newcommand{\pczm}{\left\langle S^{+}\left[S^{z},S^{-}\right] \right\rangle}
\newcommand{\ppmm}{\left\langle S^{+}S^{+}S^{-}S^{-} \right\rangle}
\newcommand{\pmpm}{\left\langle S^{+}S^{-}S^{+}S^{-} \right\rangle}
\newcommand{\pcpmm}{\left\langle S^{+}\left[S^{+},S^{-}\right]S^{-} \right\rangle}
\newcommand{\mppm}{\left\langle S^{-}S^{+}S^{+}S^{-} \right\rangle}
\newcommand{\cmppm}{\left\langle \left[S^{-},S^{+}\right]S^{+}S^{-}  \right\rangle}
\newcommand{\zpmsub}{\left\langle S^{z}\left[S\left(S+1\right)-S^{z}S^{z}+S^{z}\right] \right\rangle}
\newcommand{\zzz}{\left\langle S^{z}S^{z}S^{z}\right\rangle}
\newcommand{\zz}{\left\langle S^{z}S^{z}\right\rangle}
\newcommand{\z}{\left\langle S^{z}\right\rangle}
\newcommand{\pmsubzz}{\left\langle \left[S\left(S+1\right)-S^{z}S^{z}+S^{z}\right]S^{z}S^{z} \right\rangle}
\newcommand{\zzzz}{\left\langle S^{z}S^{z}S^{z}S^{z}\right\rangle}

Specifically, we use the results:
\begin{align*}
\pzm &=
-3 \z \zz 
+ 2\zz 
+ 2 \z^3
\\&
+\left(S\left(S+1\right)-1\right)\z 
- S\left(S+1\right), 
\\
\ppmm &=
 3 \zz^2  - 12\z\zz
\\& 
+\left(5 -2S(S+1)\right)\zz 
- 2\z^4
\\&
+ 8 \z^3
+ \left(4 S(S+1) -2\right)\z 
\\&
+S^2\left(S+1\right)^2- 2S\left(S+1\right), 
\\
\pmpm  &= 
3 \zz^2 - 6\z\zz 
\\&
+\left(1-2S\left(S+1\right)\right)\zz
- 2\z^4 
\\&
+ 4 \z^3 
+ 2 S\left(S+1\right)\z 
+ S^2\left(S+1\right)^2.
\end{align*}

\begin{widetext}

This procedure yields coupled equations of motion for the first and second moment of $S^z$ of the form:
\begin{align}
\frac{d}{dt}\left\langle S^{z}\right\rangle &=  2g^{2}\left\{ \alpha^{B}_{2} \left[S(S+1)-\ZZ-\Z\right]- \alpha^{A}_{2}\left[S(S+1) - \ZZ + \Z\right]\right\} +
\nonumber\\
&\nesp 2g^{4}\left\{\gamma^{A}_{4}\left[S(S+1)\Z -3\Z\ZZ +2\Z^3 + \ZZ \right]+\right.\nonumber\\
&\,\,\left.\gamma^{B}_{4}\left[S(S+1)\Z -3\Z\ZZ +2\Z^3 - \ZZ\right]-\right.\nonumber\\
&\,\,\left. \delta^{A}_{4} \left[S(S+1)\Z -3\Z\ZZ +2\Z^3 + 2\ZZ -S(S+1) -\Z \right]-\right.\nonumber\\
&\,\,\left. \delta^{B}_{4} \left[S(S+1)\Z -3\Z\ZZ +2\Z^3 - 2\ZZ +S(S+1) -\Z \right]\right\},
\label{cumu2}
\end{align}

\begin{align}
\frac{d}{dt}\left\langle S^{z}S^{z} \right\rangle=2g^{2}&\left\{ \alpha^{B}_{2} \left[2 S(S+1)\Z - 6\Z\ZZ + 4\Z^3  -3\ZZ + S(S + 1) - \Z  \right]-\right.\nonumber\\
&\,\, \left. \alpha^{A}_{2} \left[2 S(S+1)\Z - 6\Z\ZZ + 4\Z^3 + 3\ZZ - S(S + 1) - \Z \right]\right\}+\nonumber\\
&\nesp 4g^{4}\left\{ \alpha^{A}_{4} \left[ S^2(S + 1)^2 + 3\ZZ\ZZ - 2\Z^4 + 5\ZZ - 2S(S + 1)\ZZ \right.\right. \nonumber\\
&\,\,\left.\left. -12\Z\ZZ + 8\Z^3 + 4S(S + 1)\Z -2S(S + 1) -2\Z  \right] + \right. \nonumber\\
&\,\, \left. \alpha^{B}_{4} \left[ S^2(S + 1)^2 + 3\ZZ\ZZ - 2\Z^4 + 5\ZZ - 2S(S + 1)\ZZ \right.\right. \nonumber\\
&\,\,\left.\left. + 12\Z\ZZ -8\Z^3 - 4S(S + 1)\Z -2S(S + 1) + 2\Z  \right] \right. +\nonumber\\
&\,\, \left.\beta^{A}_{4} \left[S^2(S+1)^2 + 3\ZZ\ZZ -2\Z^4 + \ZZ -2S(S+1)\ZZ  \right.\right. \nonumber\\ 
&\,\,\left.\left. -6\Z\ZZ +4\Z^3 + 2S(S+1)\Z  \right] + \right. \nonumber\\
&\,\,\left. \beta^{B}_{4} \left[S^2(S+1)^2 + 3\ZZ\ZZ -2\Z^4 + \ZZ -2S(S+1)\ZZ  \right.\right. \nonumber\\ 
&\,\,\left.\left. +6\Z\ZZ -4\Z^3 - 2S(S+1)\Z \right] - \right. \nonumber\\
&\,\,\left. \gamma^X_{4} \left[ S^2(S+1)^2 +3\ZZ\ZZ -2\Z^4 -\ZZ 
-2S(S+1)\ZZ \right] +\right.\nonumber\\
&\,\,\left.\gamma^{A}_{4} \left[ S(S+1)\ZZ -3\ZZ\ZZ +2\Z^4 +3\Z\ZZ -2\Z^3 \right] +\right.\nonumber\\
&\,\,\left. \gamma^{B}_{4} \left[ S(S+1)\ZZ -3\ZZ\ZZ +2\Z^4 -3\Z\ZZ +2\Z^3 \right] - \right. \nonumber\\
&\,\,\left. \delta^{A}_{4} \left[ S(S+1)\ZZ -3\ZZ\ZZ +2\Z^4 +9\Z\ZZ -6\Z^3 \right.\right. \nonumber\\
&\posp\left.\left. -2S(S+1)\Z -3\ZZ +S(S+1) +\Z \right] - \right. \nonumber\\
&\,\,\left. \delta^{B}_{4} \left[S(S+1)\ZZ -3\ZZ\ZZ +2\Z^4 -9\Z\ZZ +6\Z^3 \right.\right. \nonumber\\
&\posp\left.\left. +2S(S+1)\Z -3\ZZ +S(S+1) -\Z\right]\right\}. 
\label{cumu1}
\end{align}

\end{widetext}

\newcommand{\QBp}{Q_{+}}
\newcommand{\cQBp}{Q^{\ast}_{+}}
\newcommand{\QAm}{Q_{-}}
\newcommand{\cQAm}{Q^{\ast}_{-}}

\newcommand{\Qmp}{Q_{\mp}}
\newcommand{\cQmp}{Q^{\ast}_{\mp}}

where we have defined:
\begin{align*}
    \alpha^{A(B)}_{2} &= \Qmp + \cQmp, 
    \\
    \alpha^{A(B)}_{4} &= 4\left( {\Qmp}^3 + {\cQmp}^3 \right)
    +\frac{1}{\kappa}\left( \Qmp + \cQmp \right)^2,
    \\
    \beta^{A(B)}_{4} &= 
    \frac{1}{\kappa}\left( \Qmp + \cQmp \right)^2
    -2 \Re\left[({\Qmp}^2 + \QAm\QBp)\QS \right],
    \\
    \gamma^{A(B)}_{4} &= 8 \Re\left[({\Qmp}^2 + \QAm\QBp)\QS\right],
    \\
    \gamma^X_{4} &= 
    2 \Re\left[
    ({\QAm} + {\QBp})^2\QS 
    +  2(\QAm + \cQBp)^2 \QD
    \right],
    \\
    \delta^{A(B)}_{4} &= 
    \frac{4}{\kappa}\left( \Qmp + \cQmp \right)^2.
\end{align*}

The numerical solutions of these coupled equations are shown in the Letter.  Specifically, we compare the steady state $\z$ to that found from the full equation for $\z = \sum_M M P_M$, and see a good match.  We also compare the first non-zero eigenvalue in the $k=0$ sector of the density matrix equation of motion to the eigenvalue found by linearizing Eq.~\eqref{cumu2} and Eq.~\eqref{cumu1} around the stationary states.

As noted in the Letter, Eq.~\eqref{cumu2} tends to the semiclassical limit for $N\to\infty$. In this limit, only those terms with the highest powers of $S$ or $\z$ survive (noting that $g^2 \propto 1/N$).  As a result one finds $\zz = \z^2$ in this limit (consistent with the probability distributions above), and the equation for $\z$ reduces to the semiclassical equation presented in the Letter.

\section{Comparison to the Dicke model}

\begin{figure*}[!b]
\begin{centering}
  \includegraphics[width=\textwidth]{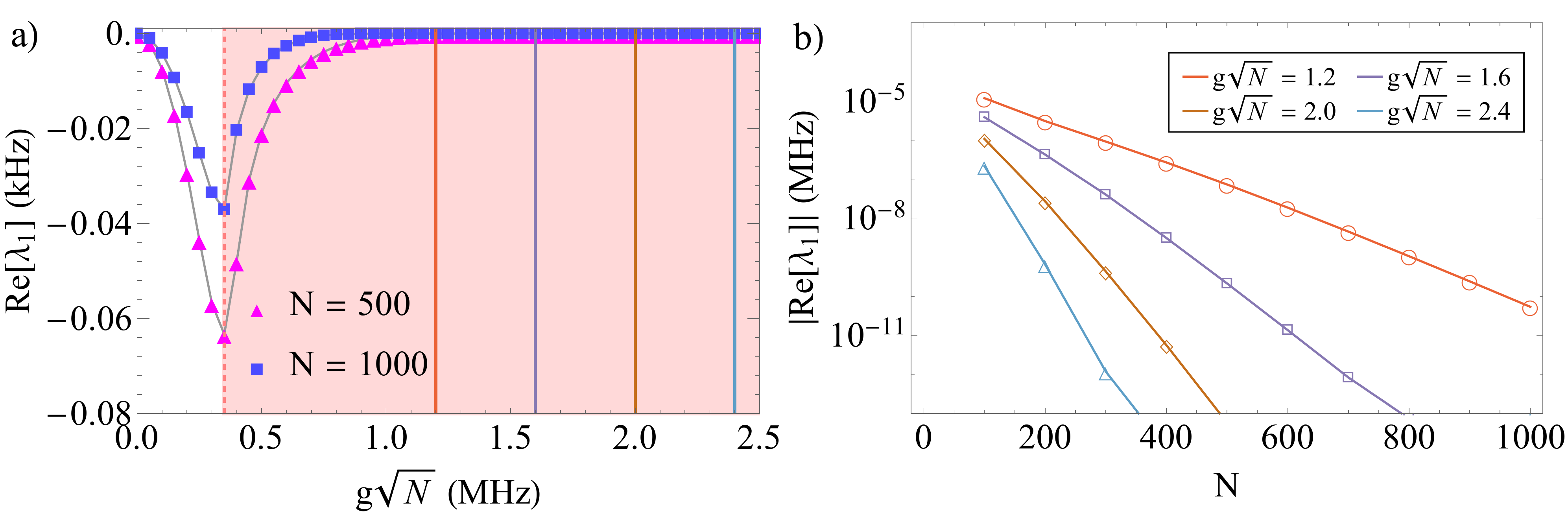}
  \par\end{centering}
  \caption{Liouvillian gap for the atom-only Redfield equation of the $\mathbb{Z}_2$ Dicke model described in Ref.~\cite{Damanet2019:AtomOnly}. (a) Shows the gap vs coupling strength at fixed $N$, while (b) shows the scaling of the gap with system size $N$ at fixed values of $g\sqrt{N}$.  The parameters used are $\kappa=8.1$ MHz, $\omega_{0}=0.01$ MHz, $\omega=5$ MHz.}
\label{figDicke}
\end{figure*}

This section compares the behavior of the Liouvillian gap in the U(1) Dicke problem, discussed in the Letter, to those for the $\mathbb{Z}_2$ Dicke model analyzed in Ref.~\cite{Damanet2019:AtomOnly}. The problem in that case is described by:
\begin{align}
\dot{\rho}_t&=-i[H,\rho_t]+\kappa\mathcal{L}[a]
\label{MESingleMode}
\\
H&=\omega_0 S^z + \omega a^\dagger a + 2g(a+a^\dagger)S^x
\label{HSingleMode}
\end{align}

In Ref.~\cite{Damanet2019:AtomOnly}, a second order Redfield equation is found to provide an adequate atom-only density matrix equation of motion, recovering the expected phase transition, and damped collective modes.  However, no analysis of the Liouvillian gap vs system size was presented; we provide this here.

In the single-mode Dicke model, because there is no U(1) symmetry, there is no separation of the density matrix equation of motion into separate sectors. Therefore, we must numerically diagonalize the complete Liouvillian.  Figure.~\ref{figDicke} shows the corresponding Liouvillian gap---Liouvillian eigenvalue with the smallest (non-zero) real part. As shown in Fig.~\ref{figDicke}(a), the gap appears to close over as one enters the superradiant phase (pink region), even  for $500$ spins.  We also show the evolution of gap with system size. It is noteworthy that the gap found this way scales differently to the U(1) problem discussed in the Letter. For the U(1) model, we found a scaling $\lambda^{(1)}_{0}=A+B/N$ while for the $\mathbb{Z}_2$ Dicke model, the behavior is  $\lambda_{1}\propto\exp(-CN)$, as shown in Fig.~\ref{figDicke}(b).


\begin{thebibliography}{64}%
\makeatletter
\providecommand \@ifxundefined [1]{%
 \@ifx{#1\undefined}
}%
\providecommand \@ifnum [1]{%
 \ifnum #1\expandafter \@firstoftwo
 \else \expandafter \@secondoftwo
 \fi
}%
\providecommand \@ifx [1]{%
 \ifx #1\expandafter \@firstoftwo
 \else \expandafter \@secondoftwo
 \fi
}%
\providecommand \natexlab [1]{#1}%
\providecommand \enquote  [1]{``#1''}%
\providecommand \bibnamefont  [1]{#1}%
\providecommand \bibfnamefont [1]{#1}%
\providecommand \citenamefont [1]{#1}%
\providecommand \href@noop [0]{\@secondoftwo}%
\providecommand \href [0]{\begingroup \@sanitize@url \@href}%
\providecommand \@href[1]{\@@startlink{#1}\@@href}%
\providecommand \@@href[1]{\endgroup#1\@@endlink}%
\providecommand \@sanitize@url [0]{\catcode `\\12\catcode `\$12\catcode
  `\&12\catcode `\#12\catcode `\^12\catcode `\_12\catcode `\%12\relax}%
\providecommand \@@startlink[1]{}%
\providecommand \@@endlink[0]{}%
\providecommand \url  [0]{\begingroup\@sanitize@url \@url }%
\providecommand \@url [1]{\endgroup\@href {#1}{\urlprefix }}%
\providecommand \urlprefix  [0]{URL }%
\providecommand \Eprint [0]{\href }%
\providecommand \doibase [0]{http://dx.doi.org/}%
\providecommand \selectlanguage [0]{\@gobble}%
\providecommand \bibinfo  [0]{\@secondoftwo}%
\providecommand \bibfield  [0]{\@secondoftwo}%
\providecommand \translation [1]{[#1]}%
\providecommand \BibitemOpen [0]{}%
\providecommand \bibitemStop [0]{}%
\providecommand \bibitemNoStop [0]{.\EOS\space}%
\providecommand \EOS [0]{\spacefactor3000\relax}%
\providecommand \BibitemShut  [1]{\csname bibitem#1\endcsname}%
\let\auto@bib@innerbib\@empty
\bibitem [{\citenamefont {Landau}\ and\ \citenamefont
  {Lifshitz}(1980)}]{landau1980statistical}%
  \BibitemOpen
  \bibfield  {author} {\bibinfo {author} {\bibfnamefont {L.}~\bibnamefont
  {Landau}}\ and\ \bibinfo {author} {\bibfnamefont {E.}~\bibnamefont
  {Lifshitz}},\ }\href@noop {} {\emph {\bibinfo {title} {Statistical
  Physics}}},\ \bibinfo {series} {Theoretical Physics}, Vol.~\bibinfo {volume}
  {5}\ (\bibinfo  {publisher} {Butterworth-Heinemann},\ \bibinfo {address}
  {Oxford},\ \bibinfo {year} {1980})\BibitemShut {NoStop}%
\bibitem [{\citenamefont {Chaikin}\ and\ \citenamefont
  {Lubensky}(1995)}]{chaikin1995principles}%
  \BibitemOpen
  \bibfield  {author} {\bibinfo {author} {\bibfnamefont {P.}~\bibnamefont
  {Chaikin}}\ and\ \bibinfo {author} {\bibfnamefont {T.}~\bibnamefont
  {Lubensky}},\ }\href@noop {} {\emph {\bibinfo {title} {Principles of
  Condensed Matter Physics}}}\ (\bibinfo  {publisher} {Cambridge University
  Press},\ \bibinfo {address} {Cambridge},\ \bibinfo {year} {1995})\BibitemShut
  {NoStop}%
\bibitem [{\citenamefont {Diehl}\ \emph {et~al.}(2008)\citenamefont {Diehl},
  \citenamefont {Micheli}, \citenamefont {Kantian}, \citenamefont {Kraus},
  \citenamefont {B{\"u}chler},\ and\ \citenamefont
  {Zoller}}]{Diehl2008:quantum}%
  \BibitemOpen
  \bibfield  {author} {\bibinfo {author} {\bibfnamefont {S.}~\bibnamefont
  {Diehl}}, \bibinfo {author} {\bibfnamefont {A.}~\bibnamefont {Micheli}},
  \bibinfo {author} {\bibfnamefont {A.}~\bibnamefont {Kantian}}, \bibinfo
  {author} {\bibfnamefont {B.}~\bibnamefont {Kraus}}, \bibinfo {author}
  {\bibfnamefont {H.}~\bibnamefont {B{\"u}chler}}, \ and\ \bibinfo {author}
  {\bibfnamefont {P.}~\bibnamefont {Zoller}},\ }\bibfield  {title} {\emph
  {\enquote {\bibinfo {title} {Quantum states and phases in driven open quantum
  systems with cold atoms},}\ }}\href {\doibase 10.1038/nphys1073} {\bibfield
  {journal} {\bibinfo  {journal} {Nat. Phys.}\ }\textbf {\bibinfo {volume}
  {4}},\ \bibinfo {pages} {878} (\bibinfo {year} {2008})}\BibitemShut {NoStop}%
\bibitem [{\citenamefont {Carusotto}\ and\ \citenamefont
  {Ciuti}(2013)}]{Carusotto2013:RMP}%
  \BibitemOpen
  \bibfield  {author} {\bibinfo {author} {\bibfnamefont {I.}~\bibnamefont
  {Carusotto}}\ and\ \bibinfo {author} {\bibfnamefont {C.}~\bibnamefont
  {Ciuti}},\ }\bibfield  {title} {\emph {\enquote {\bibinfo {title} {Quantum
  fluids of light},}\ }}\href {\doibase 10.1103/RevModPhys.85.299} {\bibfield
  {journal} {\bibinfo  {journal} {Rev. Mod. Phys.}\ }\textbf {\bibinfo {volume}
  {85}},\ \bibinfo {pages} {299} (\bibinfo {year} {2013})}\BibitemShut
  {NoStop}%
\bibitem [{\citenamefont {Le~Hur}\ \emph {et~al.}(2016)\citenamefont {Le~Hur},
  \citenamefont {Henriet}, \citenamefont {Petrescu}, \citenamefont {Plekhanov},
  \citenamefont {Roux},\ and\ \citenamefont {Schir{\'o}}}]{LeHur2016:Many}%
  \BibitemOpen
  \bibfield  {author} {\bibinfo {author} {\bibfnamefont {K.}~\bibnamefont
  {Le~Hur}}, \bibinfo {author} {\bibfnamefont {L.}~\bibnamefont {Henriet}},
  \bibinfo {author} {\bibfnamefont {A.}~\bibnamefont {Petrescu}}, \bibinfo
  {author} {\bibfnamefont {K.}~\bibnamefont {Plekhanov}}, \bibinfo {author}
  {\bibfnamefont {G.}~\bibnamefont {Roux}}, \ and\ \bibinfo {author}
  {\bibfnamefont {M.}~\bibnamefont {Schir{\'o}}},\ }\bibfield  {title} {\emph
  {\enquote {\bibinfo {title} {Many-body quantum electrodynamics networks:
  Non-equilibrium condensed matter physics with light},}\ }}\href {\doibase
  10.1016/j.crhy.2016.05.003} {\bibfield  {journal} {\bibinfo  {journal} {C. R.
  Acad. Sci.}\ }\textbf {\bibinfo {volume} {17}},\ \bibinfo {pages} {808}
  (\bibinfo {year} {2016})}\BibitemShut {NoStop}%
\bibitem [{\citenamefont {Chang}\ \emph {et~al.}(2018)\citenamefont {Chang},
  \citenamefont {Douglas}, \citenamefont {Gonz\'alez-Tudela}, \citenamefont
  {Hung},\ and\ \citenamefont {Kimble}}]{Chang2018:RMP}%
  \BibitemOpen
  \bibfield  {author} {\bibinfo {author} {\bibfnamefont {D.~E.}\ \bibnamefont
  {Chang}}, \bibinfo {author} {\bibfnamefont {J.~S.}\ \bibnamefont {Douglas}},
  \bibinfo {author} {\bibfnamefont {A.}~\bibnamefont {Gonz\'alez-Tudela}},
  \bibinfo {author} {\bibfnamefont {C.-L.}\ \bibnamefont {Hung}}, \ and\
  \bibinfo {author} {\bibfnamefont {H.~J.}\ \bibnamefont {Kimble}},\ }\bibfield
   {title} {\emph {\enquote {\bibinfo {title} {Colloquium: Quantum matter built
  from nanoscopic lattices of atoms and photons},}\ }}\href {\doibase
  10.1103/RevModPhys.90.031002} {\bibfield  {journal} {\bibinfo  {journal}
  {Rev. Mod. Phys.}\ }\textbf {\bibinfo {volume} {90}},\ \bibinfo {pages}
  {031002} (\bibinfo {year} {2018})}\BibitemShut {NoStop}%
\bibitem [{\citenamefont {Kessler}(2012)}]{Kessler2012:generalized}%
  \BibitemOpen
  \bibfield  {author} {\bibinfo {author} {\bibfnamefont {E.~M.}\ \bibnamefont
  {Kessler}},\ }\bibfield  {title} {\emph {\enquote {\bibinfo {title}
  {Generalized schrieffer-wolff formalism for dissipative systems},}\ }}\href
  {\doibase 10.1103/PhysRevA.86.012126} {\bibfield  {journal} {\bibinfo
  {journal} {Phys. Rev. A}\ }\textbf {\bibinfo {volume} {86}},\ \bibinfo
  {pages} {012126} (\bibinfo {year} {2012})}\BibitemShut {NoStop}%
\bibitem [{\citenamefont {Sciolla}\ \emph {et~al.}(2015)\citenamefont
  {Sciolla}, \citenamefont {Poletti},\ and\ \citenamefont
  {Kollath}}]{Sciolla2015:TwoTime}%
  \BibitemOpen
  \bibfield  {author} {\bibinfo {author} {\bibfnamefont {B.}~\bibnamefont
  {Sciolla}}, \bibinfo {author} {\bibfnamefont {D.}~\bibnamefont {Poletti}}, \
  and\ \bibinfo {author} {\bibfnamefont {C.}~\bibnamefont {Kollath}},\
  }\bibfield  {title} {\emph {\enquote {\bibinfo {title} {Two-time correlations
  probing the dynamics of dissipative many-body quantum systems: Aging and fast
  relaxation},}\ }}\href {\doibase 10.1103/PhysRevLett.114.170401} {\bibfield
  {journal} {\bibinfo  {journal} {Phys. Rev. Lett.}\ }\textbf {\bibinfo
  {volume} {114}},\ \bibinfo {pages} {170401} (\bibinfo {year}
  {2015})}\BibitemShut {NoStop}%
\bibitem [{\citenamefont {Lenar\v{c}i\v{c}}\ \emph {et~al.}(2018)\citenamefont
  {Lenar\v{c}i\v{c}}, \citenamefont {Lange},\ and\ \citenamefont
  {Rosch}}]{Lenarcic2018:Perturbative}%
  \BibitemOpen
  \bibfield  {author} {\bibinfo {author} {\bibfnamefont {Z.}~\bibnamefont
  {Lenar\v{c}i\v{c}}}, \bibinfo {author} {\bibfnamefont {F.}~\bibnamefont
  {Lange}}, \ and\ \bibinfo {author} {\bibfnamefont {A.}~\bibnamefont
  {Rosch}},\ }\bibfield  {title} {\emph {\enquote {\bibinfo {title}
  {Perturbative approach to weakly driven many-particle systems in the presence
  of approximate conservation laws},}\ }}\href {\doibase
  10.1103/PhysRevB.97.024302} {\bibfield  {journal} {\bibinfo  {journal} {Phys.
  Rev. B}\ }\textbf {\bibinfo {volume} {97}},\ \bibinfo {pages} {024302}
  (\bibinfo {year} {2018})}\BibitemShut {NoStop}%
\bibitem [{\citenamefont {Saideh}\ \emph {et~al.}(2020)\citenamefont {Saideh},
  \citenamefont {Finkelstein-Shapiro}, \citenamefont {No\^us}, \citenamefont
  {Pullerits},\ and\ \citenamefont {Keller}}]{Saideh2020:Projection}%
  \BibitemOpen
  \bibfield  {author} {\bibinfo {author} {\bibfnamefont {I.}~\bibnamefont
  {Saideh}}, \bibinfo {author} {\bibfnamefont {D.}~\bibnamefont
  {Finkelstein-Shapiro}}, \bibinfo {author} {\bibfnamefont {C.}~\bibnamefont
  {No\^us}}, \bibinfo {author} {\bibfnamefont {T.~o.}\ \bibnamefont
  {Pullerits}}, \ and\ \bibinfo {author} {\bibfnamefont {A.}~\bibnamefont
  {Keller}},\ }\bibfield  {title} {\emph {\enquote {\bibinfo {title}
  {Projection-based adiabatic elimination of bipartite open quantum systems},}\
  }}\href {\doibase 10.1103/PhysRevA.102.032212} {\bibfield  {journal}
  {\bibinfo  {journal} {Phys. Rev. A}\ }\textbf {\bibinfo {volume} {102}},\
  \bibinfo {pages} {032212} (\bibinfo {year} {2020})}\BibitemShut {NoStop}%
\bibitem [{\citenamefont {Redfield}(1957)}]{Redfield1957a}%
  \BibitemOpen
  \bibfield  {author} {\bibinfo {author} {\bibfnamefont {A.~G.}\ \bibnamefont
  {Redfield}},\ }\bibfield  {title} {\emph {\enquote {\bibinfo {title} {{On the
  Theory of Relaxation Processes}},}\ }}\href {\doibase 10.1147/rd.11.0019}
  {\bibfield  {journal} {\bibinfo  {journal} {IBM J. Res. Dev.}\ }\textbf
  {\bibinfo {volume} {1}},\ \bibinfo {pages} {19} (\bibinfo {year}
  {1957})}\BibitemShut {NoStop}%
\bibitem [{\citenamefont {Breuer}\ and\ \citenamefont
  {Petruccione}(2002)}]{Breuer2002}%
  \BibitemOpen
  \bibfield  {author} {\bibinfo {author} {\bibfnamefont {H.-P.}\ \bibnamefont
  {Breuer}}\ and\ \bibinfo {author} {\bibfnamefont {F.}~\bibnamefont
  {Petruccione}},\ }\href {\doibase 10.1093/acprof:oso/9780199213900.001.0001}
  {\emph {\bibinfo {title} {{The Theory of Open Quantum Systems}}}}\ (\bibinfo
  {publisher} {Oxford University Press},\ \bibinfo {address} {Oxford},\
  \bibinfo {year} {2002})\BibitemShut {NoStop}%
\bibitem [{\citenamefont {Damanet}\ \emph {et~al.}(2019)\citenamefont
  {Damanet}, \citenamefont {Daley},\ and\ \citenamefont
  {Keeling}}]{Damanet2019:AtomOnly}%
  \BibitemOpen
  \bibfield  {author} {\bibinfo {author} {\bibfnamefont {F.}~\bibnamefont
  {Damanet}}, \bibinfo {author} {\bibfnamefont {A.~J.}\ \bibnamefont {Daley}},
  \ and\ \bibinfo {author} {\bibfnamefont {J.}~\bibnamefont {Keeling}},\
  }\bibfield  {title} {\emph {\enquote {\bibinfo {title} {{Atom-only
  descriptions of the driven-dissipative Dicke model}},}\ }}\href {\doibase
  10.1103/PhysRevA.99.033845} {\bibfield  {journal} {\bibinfo  {journal} {Phys.
  Rev. A}\ }\textbf {\bibinfo {volume} {99}},\ \bibinfo {pages} {033845}
  (\bibinfo {year} {2019})}\BibitemShut {NoStop}%
\bibitem [{\citenamefont {M\"uller}\ and\ \citenamefont
  {Stace}(2017)}]{Muller2017:Deriving}%
  \BibitemOpen
  \bibfield  {author} {\bibinfo {author} {\bibfnamefont {C.}~\bibnamefont
  {M\"uller}}\ and\ \bibinfo {author} {\bibfnamefont {T.~M.}\ \bibnamefont
  {Stace}},\ }\bibfield  {title} {\emph {\enquote {\bibinfo {title} {{Deriving
  Lindblad master equations with Keldysh diagrams: Correlated gain and loss in
  higher order perturbation theory}},}\ }}\href {\doibase
  10.1103/PhysRevA.95.013847} {\bibfield  {journal} {\bibinfo  {journal} {Phys.
  Rev. A}\ }\textbf {\bibinfo {volume} {95}},\ \bibinfo {pages} {013847}
  (\bibinfo {year} {2017})}\BibitemShut {NoStop}%
\bibitem [{\citenamefont {Bloch}\ \emph {et~al.}(2008)\citenamefont {Bloch},
  \citenamefont {Dalibard},\ and\ \citenamefont {Zwerger}}]{Bloch:2008td}%
  \BibitemOpen
  \bibfield  {author} {\bibinfo {author} {\bibfnamefont {I.}~\bibnamefont
  {Bloch}}, \bibinfo {author} {\bibfnamefont {J.}~\bibnamefont {Dalibard}}, \
  and\ \bibinfo {author} {\bibfnamefont {W.}~\bibnamefont {Zwerger}},\
  }\bibfield  {title} {\emph {\enquote {\bibinfo {title} {Many-body physics
  with ultracold gases},}\ }}\href {\doibase 10.1103/RevModPhys.80.885}
  {\bibfield  {journal} {\bibinfo  {journal} {Rev. Mod. Phys.}\ }\textbf
  {\bibinfo {volume} {80}},\ \bibinfo {pages} {885} (\bibinfo {year}
  {2008})}\BibitemShut {NoStop}%
\bibitem [{\citenamefont {Domokos}\ and\ \citenamefont
  {Ritsch}(2003)}]{Ritsch03_review}%
  \BibitemOpen
  \bibfield  {author} {\bibinfo {author} {\bibfnamefont {P.}~\bibnamefont
  {Domokos}}\ and\ \bibinfo {author} {\bibfnamefont {H.}~\bibnamefont
  {Ritsch}},\ }\bibfield  {title} {\emph {\enquote {\bibinfo {title}
  {Mechanical effects of light in optical resonators},}\ }}\href {\doibase
  10.1364/JOSAB.20.001098} {\bibfield  {journal} {\bibinfo  {journal} {J. Opt.
  Soc. Am. B}\ }\textbf {\bibinfo {volume} {20}},\ \bibinfo {pages} {1098}
  (\bibinfo {year} {2003})}\BibitemShut {NoStop}%
\bibitem [{\citenamefont {Dimer}\ \emph {et~al.}(2007)\citenamefont {Dimer},
  \citenamefont {Estienne}, \citenamefont {Parkins},\ and\ \citenamefont
  {Carmichael}}]{Dimer2007:Proposed}%
  \BibitemOpen
  \bibfield  {author} {\bibinfo {author} {\bibfnamefont {F.}~\bibnamefont
  {Dimer}}, \bibinfo {author} {\bibfnamefont {B.}~\bibnamefont {Estienne}},
  \bibinfo {author} {\bibfnamefont {A.~S.}\ \bibnamefont {Parkins}}, \ and\
  \bibinfo {author} {\bibfnamefont {H.~J.}\ \bibnamefont {Carmichael}},\
  }\bibfield  {title} {\emph {\enquote {\bibinfo {title} {{Proposed realization
  of the Dicke-model quantum phase transition in an optical cavity QED
  system}},}\ }}\href {\doibase 10.1103/PhysRevA.75.013804} {\bibfield
  {journal} {\bibinfo  {journal} {Phys. Rev. A}\ }\textbf {\bibinfo {volume}
  {75}},\ \bibinfo {pages} {013804} (\bibinfo {year} {2007})}\BibitemShut
  {NoStop}%
\bibitem [{\citenamefont {Baumann}\ \emph {et~al.}(2010)\citenamefont
  {Baumann}, \citenamefont {Guerlin}, \citenamefont {Brennecke},\ and\
  \citenamefont {Esslinger}}]{Baumann2010}%
  \BibitemOpen
  \bibfield  {author} {\bibinfo {author} {\bibfnamefont {K.}~\bibnamefont
  {Baumann}}, \bibinfo {author} {\bibfnamefont {C.}~\bibnamefont {Guerlin}},
  \bibinfo {author} {\bibfnamefont {F.}~\bibnamefont {Brennecke}}, \ and\
  \bibinfo {author} {\bibfnamefont {T.}~\bibnamefont {Esslinger}},\ }\bibfield
  {title} {\emph {\enquote {\bibinfo {title} {{Dicke quantum phase transition
  with a superfluid gas in an optical cavity}},}\ }}\href {\doibase
  10.1038/nature09009} {\bibfield  {journal} {\bibinfo  {journal} {Nature}\
  }\textbf {\bibinfo {volume} {464}},\ \bibinfo {pages} {1301} (\bibinfo {year}
  {2010})}\BibitemShut {NoStop}%
\bibitem [{\citenamefont {Ke\ss{}ler}\ \emph {et~al.}(2014)\citenamefont
  {Ke\ss{}ler}, \citenamefont {Klinder}, \citenamefont {Wolke},\ and\
  \citenamefont {Hemmerich}}]{Hemmerich14}%
  \BibitemOpen
  \bibfield  {author} {\bibinfo {author} {\bibfnamefont {H.}~\bibnamefont
  {Ke\ss{}ler}}, \bibinfo {author} {\bibfnamefont {J.}~\bibnamefont {Klinder}},
  \bibinfo {author} {\bibfnamefont {M.}~\bibnamefont {Wolke}}, \ and\ \bibinfo
  {author} {\bibfnamefont {A.}~\bibnamefont {Hemmerich}},\ }\bibfield  {title}
  {\emph {\enquote {\bibinfo {title} {Steering matter wave superradiance with
  an ultranarrow-band optical cavity},}\ }}\href {\doibase
  10.1103/PhysRevLett.113.070404} {\bibfield  {journal} {\bibinfo  {journal}
  {Phys. Rev. Lett.}\ }\textbf {\bibinfo {volume} {113}},\ \bibinfo {pages}
  {070404} (\bibinfo {year} {2014})}\BibitemShut {NoStop}%
\bibitem [{\citenamefont {Klinder}\ \emph {et~al.}(2015)\citenamefont
  {Klinder}, \citenamefont {Ke{\ss}ler}, \citenamefont {Bakhtiari},
  \citenamefont {Thorwart},\ and\ \citenamefont {Hemmerich}}]{Klinder2015}%
  \BibitemOpen
  \bibfield  {author} {\bibinfo {author} {\bibfnamefont {J.}~\bibnamefont
  {Klinder}}, \bibinfo {author} {\bibfnamefont {H.}~\bibnamefont {Ke{\ss}ler}},
  \bibinfo {author} {\bibfnamefont {M.~R.}\ \bibnamefont {Bakhtiari}}, \bibinfo
  {author} {\bibfnamefont {M.}~\bibnamefont {Thorwart}}, \ and\ \bibinfo
  {author} {\bibfnamefont {A.}~\bibnamefont {Hemmerich}},\ }\bibfield  {title}
  {\emph {\enquote {\bibinfo {title} {{Observation of a superradiant Mott
  insulator in the Dicke-Hubbard model}},}\ }}\href {\doibase
  10.1103/PhysRevLett.115.230403} {\bibfield  {journal} {\bibinfo  {journal}
  {Phys. Rev. Lett.}\ }\textbf {\bibinfo {volume} {115}},\ \bibinfo {pages}
  {230403} (\bibinfo {year} {2015})}\BibitemShut {NoStop}%
\bibitem [{\citenamefont {Landig}\ \emph {et~al.}(2016)\citenamefont {Landig},
  \citenamefont {Hruby}, \citenamefont {Dogra}, \citenamefont {Landini},
  \citenamefont {Mottl}, \citenamefont {Donner},\ and\ \citenamefont
  {Esslinger}}]{Landig2016}%
  \BibitemOpen
  \bibfield  {author} {\bibinfo {author} {\bibfnamefont {R.}~\bibnamefont
  {Landig}}, \bibinfo {author} {\bibfnamefont {L.}~\bibnamefont {Hruby}},
  \bibinfo {author} {\bibfnamefont {N.}~\bibnamefont {Dogra}}, \bibinfo
  {author} {\bibfnamefont {M.}~\bibnamefont {Landini}}, \bibinfo {author}
  {\bibfnamefont {R.}~\bibnamefont {Mottl}}, \bibinfo {author} {\bibfnamefont
  {T.}~\bibnamefont {Donner}}, \ and\ \bibinfo {author} {\bibfnamefont
  {T.}~\bibnamefont {Esslinger}},\ }\bibfield  {title} {\emph {\enquote
  {\bibinfo {title} {{Quantum phases from competing short- and long-range
  interactions in an optical lattice}},}\ }}\href
  {https://doi.org/10.1038/nature17409} {\bibfield  {journal} {\bibinfo
  {journal} {Nature}\ }\textbf {\bibinfo {volume} {532}},\ \bibinfo {pages}
  {476} (\bibinfo {year} {2016})}\BibitemShut {NoStop}%
\bibitem [{\citenamefont {Koll{\'{a}}r}\ \emph {et~al.}(2017)\citenamefont
  {Koll{\'{a}}r}, \citenamefont {Papageorge}, \citenamefont {Vaidya},
  \citenamefont {Guo}, \citenamefont {Keeling},\ and\ \citenamefont
  {Lev}}]{Kollar2017sm}%
  \BibitemOpen
  \bibfield  {author} {\bibinfo {author} {\bibfnamefont {A.~J.}\ \bibnamefont
  {Koll{\'{a}}r}}, \bibinfo {author} {\bibfnamefont {A.~T.}\ \bibnamefont
  {Papageorge}}, \bibinfo {author} {\bibfnamefont {V.~D.}\ \bibnamefont
  {Vaidya}}, \bibinfo {author} {\bibfnamefont {Y.}~\bibnamefont {Guo}},
  \bibinfo {author} {\bibfnamefont {J.}~\bibnamefont {Keeling}}, \ and\
  \bibinfo {author} {\bibfnamefont {B.~L.}\ \bibnamefont {Lev}},\ }\bibfield
  {title} {\emph {\enquote {\bibinfo {title} {{Supermode-density-wave-polariton
  condensation with a Bose-Einstein condensate in a multimode cavity}},}\
  }}\href {\doibase 10.1038/ncomms14386} {\bibfield  {journal} {\bibinfo
  {journal} {Nat. Commun.}\ }\textbf {\bibinfo {volume} {8}},\ \bibinfo {pages}
  {14386} (\bibinfo {year} {2017})}\BibitemShut {NoStop}%
\bibitem [{\citenamefont {L{\'e}onard}\ \emph
  {et~al.}(2017{\natexlab{a}})\citenamefont {L{\'e}onard}, \citenamefont
  {Morales}, \citenamefont {Zupancic}, \citenamefont {Esslinger},\ and\
  \citenamefont {Donner}}]{Leonard2017:Supersolid}%
  \BibitemOpen
  \bibfield  {author} {\bibinfo {author} {\bibfnamefont {J.}~\bibnamefont
  {L{\'e}onard}}, \bibinfo {author} {\bibfnamefont {A.}~\bibnamefont
  {Morales}}, \bibinfo {author} {\bibfnamefont {P.}~\bibnamefont {Zupancic}},
  \bibinfo {author} {\bibfnamefont {T.}~\bibnamefont {Esslinger}}, \ and\
  \bibinfo {author} {\bibfnamefont {T.}~\bibnamefont {Donner}},\ }\bibfield
  {title} {\emph {\enquote {\bibinfo {title} {{Supersolid formation in a
  quantum gas breaking a continuous translational symmetry}},}\ }}\href
  {\doibase 10.1038/nature21067} {\bibfield  {journal} {\bibinfo  {journal}
  {Nature}\ }\textbf {\bibinfo {volume} {543}},\ \bibinfo {pages} {87}
  (\bibinfo {year} {2017}{\natexlab{a}})}\BibitemShut {NoStop}%
\bibitem [{\citenamefont {Vaidya}\ \emph {et~al.}(2018)\citenamefont {Vaidya},
  \citenamefont {Guo}, \citenamefont {Kroeze}, \citenamefont {Ballantine},
  \citenamefont {Koll\'ar}, \citenamefont {Keeling},\ and\ \citenamefont
  {Lev}}]{vaidya2018tunable}%
  \BibitemOpen
  \bibfield  {author} {\bibinfo {author} {\bibfnamefont {V.~D.}\ \bibnamefont
  {Vaidya}}, \bibinfo {author} {\bibfnamefont {Y.}~\bibnamefont {Guo}},
  \bibinfo {author} {\bibfnamefont {R.~M.}\ \bibnamefont {Kroeze}}, \bibinfo
  {author} {\bibfnamefont {K.~E.}\ \bibnamefont {Ballantine}}, \bibinfo
  {author} {\bibfnamefont {A.~J.}\ \bibnamefont {Koll\'ar}}, \bibinfo {author}
  {\bibfnamefont {J.}~\bibnamefont {Keeling}}, \ and\ \bibinfo {author}
  {\bibfnamefont {B.~L.}\ \bibnamefont {Lev}},\ }\bibfield  {title} {\emph
  {\enquote {\bibinfo {title} {Tunable-range, photon-mediated atomic
  interactions in multimode cavity qed},}\ }}\href {\doibase
  10.1103/PhysRevX.8.011002} {\bibfield  {journal} {\bibinfo  {journal} {Phys.
  Rev. X}\ }\textbf {\bibinfo {volume} {8}},\ \bibinfo {pages} {011002}
  (\bibinfo {year} {2018})}\BibitemShut {NoStop}%
\bibitem [{\citenamefont {Landini}\ \emph {et~al.}(2018)\citenamefont
  {Landini}, \citenamefont {Dogra}, \citenamefont {Kroeger}, \citenamefont
  {Hruby}, \citenamefont {Donner},\ and\ \citenamefont
  {Esslinger}}]{Landini2018:Formation}%
  \BibitemOpen
  \bibfield  {author} {\bibinfo {author} {\bibfnamefont {M.}~\bibnamefont
  {Landini}}, \bibinfo {author} {\bibfnamefont {N.}~\bibnamefont {Dogra}},
  \bibinfo {author} {\bibfnamefont {K.}~\bibnamefont {Kroeger}}, \bibinfo
  {author} {\bibfnamefont {L.}~\bibnamefont {Hruby}}, \bibinfo {author}
  {\bibfnamefont {T.}~\bibnamefont {Donner}}, \ and\ \bibinfo {author}
  {\bibfnamefont {T.}~\bibnamefont {Esslinger}},\ }\bibfield  {title} {\emph
  {\enquote {\bibinfo {title} {Formation of a spin texture in a quantum gas
  coupled to a cavity},}\ }}\href {\doibase 10.1103/PhysRevLett.120.223602}
  {\bibfield  {journal} {\bibinfo  {journal} {Phys. Rev. Lett.}\ }\textbf
  {\bibinfo {volume} {120}},\ \bibinfo {pages} {223602} (\bibinfo {year}
  {2018})}\BibitemShut {NoStop}%
\bibitem [{\citenamefont {Kroeze}\ \emph {et~al.}(2018)\citenamefont {Kroeze},
  \citenamefont {Guo}, \citenamefont {Vaidya}, \citenamefont {Keeling},\ and\
  \citenamefont {Lev}}]{Kroeze2018:Spinor}%
  \BibitemOpen
  \bibfield  {author} {\bibinfo {author} {\bibfnamefont {R.~M.}\ \bibnamefont
  {Kroeze}}, \bibinfo {author} {\bibfnamefont {Y.}~\bibnamefont {Guo}},
  \bibinfo {author} {\bibfnamefont {V.~D.}\ \bibnamefont {Vaidya}}, \bibinfo
  {author} {\bibfnamefont {J.}~\bibnamefont {Keeling}}, \ and\ \bibinfo
  {author} {\bibfnamefont {B.~L.}\ \bibnamefont {Lev}},\ }\bibfield  {title}
  {\emph {\enquote {\bibinfo {title} {Spinor self-ordering of a quantum gas in
  a cavity},}\ }}\href {\doibase 10.1103/PhysRevLett.121.163601} {\bibfield
  {journal} {\bibinfo  {journal} {Phys. Rev. Lett.}\ }\textbf {\bibinfo
  {volume} {121}},\ \bibinfo {pages} {163601} (\bibinfo {year}
  {2018})}\BibitemShut {NoStop}%
\bibitem [{\citenamefont {Zhang}\ \emph {et~al.}(2018)\citenamefont {Zhang},
  \citenamefont {Lee}, \citenamefont {Kumar}, \citenamefont {Arnold},
  \citenamefont {Masson}, \citenamefont {Grimsmo}, \citenamefont {Parkins},\
  and\ \citenamefont {Barrett}}]{Zhang2018:Dicke}%
  \BibitemOpen
  \bibfield  {author} {\bibinfo {author} {\bibfnamefont {Z.}~\bibnamefont
  {Zhang}}, \bibinfo {author} {\bibfnamefont {C.~H.}\ \bibnamefont {Lee}},
  \bibinfo {author} {\bibfnamefont {R.}~\bibnamefont {Kumar}}, \bibinfo
  {author} {\bibfnamefont {K.~J.}\ \bibnamefont {Arnold}}, \bibinfo {author}
  {\bibfnamefont {S.~J.}\ \bibnamefont {Masson}}, \bibinfo {author}
  {\bibfnamefont {A.~L.}\ \bibnamefont {Grimsmo}}, \bibinfo {author}
  {\bibfnamefont {A.~S.}\ \bibnamefont {Parkins}}, \ and\ \bibinfo {author}
  {\bibfnamefont {M.~D.}\ \bibnamefont {Barrett}},\ }\bibfield  {title} {\emph
  {\enquote {\bibinfo {title} {Dicke-model simulation via cavity-assisted raman
  transitions},}\ }}\href {\doibase 10.1103/PhysRevA.97.043858} {\bibfield
  {journal} {\bibinfo  {journal} {Phys. Rev. A}\ }\textbf {\bibinfo {volume}
  {97}},\ \bibinfo {pages} {043858} (\bibinfo {year} {2018})}\BibitemShut
  {NoStop}%
\bibitem [{\citenamefont {Guo}\ \emph {et~al.}(2019{\natexlab{a}})\citenamefont
  {Guo}, \citenamefont {Kroeze}, \citenamefont {Vaidya}, \citenamefont
  {Keeling},\ and\ \citenamefont {Lev}}]{Guo2019Sign}%
  \BibitemOpen
  \bibfield  {author} {\bibinfo {author} {\bibfnamefont {Y.}~\bibnamefont
  {Guo}}, \bibinfo {author} {\bibfnamefont {R.~M.}\ \bibnamefont {Kroeze}},
  \bibinfo {author} {\bibfnamefont {V.~D.}\ \bibnamefont {Vaidya}}, \bibinfo
  {author} {\bibfnamefont {J.}~\bibnamefont {Keeling}}, \ and\ \bibinfo
  {author} {\bibfnamefont {B.~L.}\ \bibnamefont {Lev}},\ }\bibfield  {title}
  {\emph {\enquote {\bibinfo {title} {Sign-changing photon-mediated atom
  interactions in multimode cavity quantum electrodynamics},}\ }}\href
  {\doibase 10.1103/PhysRevLett.122.193601} {\bibfield  {journal} {\bibinfo
  {journal} {Phys. Rev. Lett.}\ }\textbf {\bibinfo {volume} {122}},\ \bibinfo
  {pages} {193601} (\bibinfo {year} {2019}{\natexlab{a}})}\BibitemShut
  {NoStop}%
\bibitem [{\citenamefont {Kroeze}\ \emph {et~al.}(2019)\citenamefont {Kroeze},
  \citenamefont {Guo},\ and\ \citenamefont {Lev}}]{Kroeze2019:Dynamical}%
  \BibitemOpen
  \bibfield  {author} {\bibinfo {author} {\bibfnamefont {R.~M.}\ \bibnamefont
  {Kroeze}}, \bibinfo {author} {\bibfnamefont {Y.}~\bibnamefont {Guo}}, \ and\
  \bibinfo {author} {\bibfnamefont {B.~L.}\ \bibnamefont {Lev}},\ }\bibfield
  {title} {\emph {\enquote {\bibinfo {title} {Dynamical spin-orbit coupling of
  a quantum gas},}\ }}\href {\doibase 10.1103/PhysRevLett.123.160404}
  {\bibfield  {journal} {\bibinfo  {journal} {Phys. Rev. Lett.}\ }\textbf
  {\bibinfo {volume} {123}},\ \bibinfo {pages} {160404} (\bibinfo {year}
  {2019})}\BibitemShut {NoStop}%
\bibitem [{\citenamefont {Davis}\ \emph {et~al.}(2019)\citenamefont {Davis},
  \citenamefont {Bentsen}, \citenamefont {Homeier}, \citenamefont {Li},\ and\
  \citenamefont {Schleier-Smith}}]{Davis2019:Photon}%
  \BibitemOpen
  \bibfield  {author} {\bibinfo {author} {\bibfnamefont {E.~J.}\ \bibnamefont
  {Davis}}, \bibinfo {author} {\bibfnamefont {G.}~\bibnamefont {Bentsen}},
  \bibinfo {author} {\bibfnamefont {L.}~\bibnamefont {Homeier}}, \bibinfo
  {author} {\bibfnamefont {T.}~\bibnamefont {Li}}, \ and\ \bibinfo {author}
  {\bibfnamefont {M.~H.}\ \bibnamefont {Schleier-Smith}},\ }\bibfield  {title}
  {\emph {\enquote {\bibinfo {title} {Photon-mediated spin-exchange dynamics of
  spin-1 atoms},}\ }}\href {\doibase 10.1103/PhysRevLett.122.010405} {\bibfield
   {journal} {\bibinfo  {journal} {Phys. Rev. Lett.}\ }\textbf {\bibinfo
  {volume} {122}},\ \bibinfo {pages} {010405} (\bibinfo {year}
  {2019})}\BibitemShut {NoStop}%
\bibitem [{\citenamefont {Hepp}\ and\ \citenamefont
  {Lieb}(1973)}]{hepp1973equilibrium}%
  \BibitemOpen
  \bibfield  {author} {\bibinfo {author} {\bibfnamefont {K.}~\bibnamefont
  {Hepp}}\ and\ \bibinfo {author} {\bibfnamefont {E.~H.}\ \bibnamefont
  {Lieb}},\ }\bibfield  {title} {\emph {\enquote {\bibinfo {title} {Equilibrium
  statistical mechanics of matter interacting with the quantized radiation
  field},}\ }}\href {\doibase 10.1103/PhysRevA.8.2517} {\bibfield  {journal}
  {\bibinfo  {journal} {Phys. Rev. A}\ }\textbf {\bibinfo {volume} {8}},\
  \bibinfo {pages} {2517} (\bibinfo {year} {1973})}\BibitemShut {NoStop}%
\bibitem [{\citenamefont {Wang}\ and\ \citenamefont {Hioe}(1973)}]{wang73}%
  \BibitemOpen
  \bibfield  {author} {\bibinfo {author} {\bibfnamefont {Y.~K.}\ \bibnamefont
  {Wang}}\ and\ \bibinfo {author} {\bibfnamefont {F.~T.}\ \bibnamefont
  {Hioe}},\ }\bibfield  {title} {\emph {\enquote {\bibinfo {title} {Phase
  transition in the dicke model of superradiance},}\ }}\href {\doibase
  10.1103/PhysRevA.7.831} {\bibfield  {journal} {\bibinfo  {journal} {Phys.
  Rev. A}\ }\textbf {\bibinfo {volume} {7}},\ \bibinfo {pages} {831} (\bibinfo
  {year} {1973})}\BibitemShut {NoStop}%
\bibitem [{\citenamefont {Garraway}(2011)}]{Garraway2011}%
  \BibitemOpen
  \bibfield  {author} {\bibinfo {author} {\bibfnamefont {B.~M.}\ \bibnamefont
  {Garraway}},\ }\bibfield  {title} {\emph {\enquote {\bibinfo {title} {{The
  Dicke model in quantum optics: Dicke model revisited}},}\ }}\href
  {https://doi.org/10.1098/rsta.2010.0333} {\bibfield  {journal} {\bibinfo
  {journal} {Philos. Trans. R. Soc. London, Ser. A}\ }\textbf {\bibinfo
  {volume} {369}},\ \bibinfo {pages} {1137} (\bibinfo {year}
  {2011})}\BibitemShut {NoStop}%
\bibitem [{\citenamefont {Kirton}\ \emph {et~al.}(2019)\citenamefont {Kirton},
  \citenamefont {Roses}, \citenamefont {Keeling},\ and\ \citenamefont
  {Dalla~Torre}}]{kirton2019:Review}%
  \BibitemOpen
  \bibfield  {author} {\bibinfo {author} {\bibfnamefont {P.}~\bibnamefont
  {Kirton}}, \bibinfo {author} {\bibfnamefont {M.~M.}\ \bibnamefont {Roses}},
  \bibinfo {author} {\bibfnamefont {J.}~\bibnamefont {Keeling}}, \ and\
  \bibinfo {author} {\bibfnamefont {E.~G.}\ \bibnamefont {Dalla~Torre}},\
  }\bibfield  {title} {\emph {\enquote {\bibinfo {title} {{Introduction to the
  Dicke Model: From Equilibrium to Nonequilibrium, and Vice Versa}},}\ }}\href
  {\doibase 10.1002/qute.201800043} {\bibfield  {journal} {\bibinfo  {journal}
  {Adv. Quantum Technol.}\ }\textbf {\bibinfo {volume} {2}},\ \bibinfo {pages}
  {1800043} (\bibinfo {year} {2019})}\BibitemShut {NoStop}%
\bibitem [{\citenamefont {Fan}\ \emph {et~al.}(2014)\citenamefont {Fan},
  \citenamefont {Yang}, \citenamefont {Zhang}, \citenamefont {Ma},
  \citenamefont {Chen},\ and\ \citenamefont {Jia}}]{Fan2014:Hidden}%
  \BibitemOpen
  \bibfield  {author} {\bibinfo {author} {\bibfnamefont {J.}~\bibnamefont
  {Fan}}, \bibinfo {author} {\bibfnamefont {Z.}~\bibnamefont {Yang}}, \bibinfo
  {author} {\bibfnamefont {Y.}~\bibnamefont {Zhang}}, \bibinfo {author}
  {\bibfnamefont {J.}~\bibnamefont {Ma}}, \bibinfo {author} {\bibfnamefont
  {G.}~\bibnamefont {Chen}}, \ and\ \bibinfo {author} {\bibfnamefont
  {S.}~\bibnamefont {Jia}},\ }\bibfield  {title} {\emph {\enquote {\bibinfo
  {title} {Hidden continuous symmetry and nambu-goldstone mode in a two-mode
  dicke model},}\ }}\href {\doibase 10.1103/PhysRevA.89.023812} {\bibfield
  {journal} {\bibinfo  {journal} {Phys. Rev. A}\ }\textbf {\bibinfo {volume}
  {89}},\ \bibinfo {pages} {023812} (\bibinfo {year} {2014})}\BibitemShut
  {NoStop}%
\bibitem [{\citenamefont {Baksic}\ and\ \citenamefont
  {Ciuti}(2014)}]{baksic2014controlling}%
  \BibitemOpen
  \bibfield  {author} {\bibinfo {author} {\bibfnamefont {A.}~\bibnamefont
  {Baksic}}\ and\ \bibinfo {author} {\bibfnamefont {C.}~\bibnamefont {Ciuti}},\
  }\bibfield  {title} {\emph {\enquote {\bibinfo {title} {Controlling discrete
  and continuous symmetries in ``superradiant'' phase transitions with circuit
  qed systems},}\ }}\href {\doibase 10.1103/PhysRevLett.112.173601} {\bibfield
  {journal} {\bibinfo  {journal} {Phys. Rev. Lett.}\ }\textbf {\bibinfo
  {volume} {112}},\ \bibinfo {pages} {173601} (\bibinfo {year}
  {2014})}\BibitemShut {NoStop}%
\bibitem [{\citenamefont {L{\'e}onard}\ \emph
  {et~al.}(2017{\natexlab{b}})\citenamefont {L{\'e}onard}, \citenamefont
  {Morales}, \citenamefont {Zupancic}, \citenamefont {Donner},\ and\
  \citenamefont {Esslinger}}]{Leonard2017:Monitoring}%
  \BibitemOpen
  \bibfield  {author} {\bibinfo {author} {\bibfnamefont {J.}~\bibnamefont
  {L{\'e}onard}}, \bibinfo {author} {\bibfnamefont {A.}~\bibnamefont
  {Morales}}, \bibinfo {author} {\bibfnamefont {P.}~\bibnamefont {Zupancic}},
  \bibinfo {author} {\bibfnamefont {T.}~\bibnamefont {Donner}}, \ and\ \bibinfo
  {author} {\bibfnamefont {T.}~\bibnamefont {Esslinger}},\ }\bibfield  {title}
  {\emph {\enquote {\bibinfo {title} {Monitoring and manipulating higgs and
  goldstone modes in a supersolid quantum gas},}\ }}\href {\doibase
  10.1126/science.aan2608} {\bibfield  {journal} {\bibinfo  {journal}
  {Science}\ }\textbf {\bibinfo {volume} {358}},\ \bibinfo {pages} {1415}
  (\bibinfo {year} {2017}{\natexlab{b}})}\BibitemShut {NoStop}%
\bibitem [{\citenamefont {Moodie}\ \emph {et~al.}(2018)\citenamefont {Moodie},
  \citenamefont {Ballantine},\ and\ \citenamefont
  {Keeling}}]{Moodie2018:Generalized}%
  \BibitemOpen
  \bibfield  {author} {\bibinfo {author} {\bibfnamefont {R.~I.}\ \bibnamefont
  {Moodie}}, \bibinfo {author} {\bibfnamefont {K.~E.}\ \bibnamefont
  {Ballantine}}, \ and\ \bibinfo {author} {\bibfnamefont {J.}~\bibnamefont
  {Keeling}},\ }\bibfield  {title} {\emph {\enquote {\bibinfo {title}
  {{Generalized classes of continuous symmetries in two-mode Dicke models}},}\
  }}\href {\doibase 10.1103/PhysRevA.97.033802} {\bibfield  {journal} {\bibinfo
   {journal} {Phys. Rev. A}\ }\textbf {\bibinfo {volume} {97}},\ \bibinfo
  {pages} {033802} (\bibinfo {year} {2018})}\BibitemShut {NoStop}%
\bibitem [{\citenamefont {Chiacchio}\ and\ \citenamefont
  {Nunnenkamp}(2018)}]{Chiacchio2018:Emergence}%
  \BibitemOpen
  \bibfield  {author} {\bibinfo {author} {\bibfnamefont {E.~I.~R.}\
  \bibnamefont {Chiacchio}}\ and\ \bibinfo {author} {\bibfnamefont
  {A.}~\bibnamefont {Nunnenkamp}},\ }\bibfield  {title} {\emph {\enquote
  {\bibinfo {title} {Emergence of continuous rotational symmetries in ultracold
  atoms coupled to optical cavities},}\ }}\href {\doibase
  10.1103/PhysRevA.98.023617} {\bibfield  {journal} {\bibinfo  {journal} {Phys.
  Rev. A}\ }\textbf {\bibinfo {volume} {98}},\ \bibinfo {pages} {023617}
  (\bibinfo {year} {2018})}\BibitemShut {NoStop}%
\bibitem [{\citenamefont {Gopalakrishnan}\ \emph {et~al.}(2009)\citenamefont
  {Gopalakrishnan}, \citenamefont {Lev},\ and\ \citenamefont
  {Goldbart}}]{Gopalakrishnan09}%
  \BibitemOpen
  \bibfield  {author} {\bibinfo {author} {\bibfnamefont {S.}~\bibnamefont
  {Gopalakrishnan}}, \bibinfo {author} {\bibfnamefont {B.~L.}\ \bibnamefont
  {Lev}}, \ and\ \bibinfo {author} {\bibfnamefont {P.~M.}\ \bibnamefont
  {Goldbart}},\ }\bibfield  {title} {\emph {\enquote {\bibinfo {title}
  {Emergent crystallinity and frustration with bose--einstein condensates in
  multimode cavities},}\ }}\href {\doibase 10.1038/nphys1403} {\bibfield
  {journal} {\bibinfo  {journal} {Nat. Phys.}\ }\textbf {\bibinfo {volume}
  {5}},\ \bibinfo {pages} {845} (\bibinfo {year} {2009})}\BibitemShut {NoStop}%
\bibitem [{\citenamefont {Gopalakrishnan}\ \emph {et~al.}(2010)\citenamefont
  {Gopalakrishnan}, \citenamefont {Lev},\ and\ \citenamefont
  {Goldbart}}]{Gopalakrishnan10}%
  \BibitemOpen
  \bibfield  {author} {\bibinfo {author} {\bibfnamefont {S.}~\bibnamefont
  {Gopalakrishnan}}, \bibinfo {author} {\bibfnamefont {B.~L.}\ \bibnamefont
  {Lev}}, \ and\ \bibinfo {author} {\bibfnamefont {P.~M.}\ \bibnamefont
  {Goldbart}},\ }\bibfield  {title} {\emph {\enquote {\bibinfo {title}
  {{Atom-light crystallization of Bose-Einstein condensates in multimode
  cavities: Nonequilibrium classical and quantum phase transitions, emergent
  lattices, supersolidity, and frustration}},}\ }}\href
  {https://doi.org/10.1103/PhysRevA.82.043612} {\bibfield  {journal} {\bibinfo
  {journal} {Phys. Rev. A}\ }\textbf {\bibinfo {volume} {82}},\ \bibinfo
  {pages} {043612} (\bibinfo {year} {2010})}\BibitemShut {NoStop}%
\bibitem [{\citenamefont {Labeyrie}\ \emph {et~al.}(2014)\citenamefont
  {Labeyrie}, \citenamefont {Tesio}, \citenamefont {Gomes}, \citenamefont
  {Oppo}, \citenamefont {Firth}, \citenamefont {Robb}, \citenamefont {Arnold},
  \citenamefont {Kaiser},\ and\ \citenamefont {Ackemann}}]{Labeyrie:2014gh}%
  \BibitemOpen
  \bibfield  {author} {\bibinfo {author} {\bibfnamefont {G.}~\bibnamefont
  {Labeyrie}}, \bibinfo {author} {\bibfnamefont {E.}~\bibnamefont {Tesio}},
  \bibinfo {author} {\bibfnamefont {P.~M.}\ \bibnamefont {Gomes}}, \bibinfo
  {author} {\bibfnamefont {G.~L.}\ \bibnamefont {Oppo}}, \bibinfo {author}
  {\bibfnamefont {W.~J.}\ \bibnamefont {Firth}}, \bibinfo {author}
  {\bibfnamefont {G.~R.~M.}\ \bibnamefont {Robb}}, \bibinfo {author}
  {\bibfnamefont {A.~S.}\ \bibnamefont {Arnold}}, \bibinfo {author}
  {\bibfnamefont {R.}~\bibnamefont {Kaiser}}, \ and\ \bibinfo {author}
  {\bibfnamefont {T.}~\bibnamefont {Ackemann}},\ }\bibfield  {title} {\emph
  {\enquote {\bibinfo {title} {{Optomechanical self-structuring in a cold
  atomic gas}},}\ }}\href {\doibase 10.1038/nphoton.2014.52} {\bibfield
  {journal} {\bibinfo  {journal} {Nat. Photonics}\ }\textbf {\bibinfo {volume}
  {8}},\ \bibinfo {pages} {321} (\bibinfo {year} {2014})}\BibitemShut {NoStop}%
\bibitem [{\citenamefont {Ballantine}\ \emph {et~al.}(2017)\citenamefont
  {Ballantine}, \citenamefont {Lev},\ and\ \citenamefont
  {Keeling}}]{ballantine2017meissner}%
  \BibitemOpen
  \bibfield  {author} {\bibinfo {author} {\bibfnamefont {K.~E.}\ \bibnamefont
  {Ballantine}}, \bibinfo {author} {\bibfnamefont {B.~L.}\ \bibnamefont {Lev}},
  \ and\ \bibinfo {author} {\bibfnamefont {J.}~\bibnamefont {Keeling}},\
  }\bibfield  {title} {\emph {\enquote {\bibinfo {title} {Meissner-like effect
  for a synthetic gauge field in multimode cavity qed},}\ }}\href {\doibase
  10.1103/PhysRevLett.118.045302} {\bibfield  {journal} {\bibinfo  {journal}
  {Phys. Rev. Lett.}\ }\textbf {\bibinfo {volume} {118}},\ \bibinfo {pages}
  {045302} (\bibinfo {year} {2017})}\BibitemShut {NoStop}%
\bibitem [{\citenamefont {Rylands}\ \emph {et~al.}(2020)\citenamefont
  {Rylands}, \citenamefont {Guo}, \citenamefont {Lev}, \citenamefont
  {Keeling},\ and\ \citenamefont {Galitski}}]{rylands2020photon}%
  \BibitemOpen
  \bibfield  {author} {\bibinfo {author} {\bibfnamefont {C.}~\bibnamefont
  {Rylands}}, \bibinfo {author} {\bibfnamefont {Y.}~\bibnamefont {Guo}},
  \bibinfo {author} {\bibfnamefont {B.~L.}\ \bibnamefont {Lev}}, \bibinfo
  {author} {\bibfnamefont {J.}~\bibnamefont {Keeling}}, \ and\ \bibinfo
  {author} {\bibfnamefont {V.}~\bibnamefont {Galitski}},\ }\bibfield  {title}
  {\emph {\enquote {\bibinfo {title} {Photon-mediated peierls transition of a
  1d gas in a multimode optical cavity},}\ }}\href {\doibase
  10.1103/PhysRevLett.125.010404} {\bibfield  {journal} {\bibinfo  {journal}
  {Phys. Rev. Lett.}\ }\textbf {\bibinfo {volume} {125}},\ \bibinfo {pages}
  {010404} (\bibinfo {year} {2020})}\BibitemShut {NoStop}%
\bibitem [{\citenamefont {Guo}\ \emph {et~al.}(2019{\natexlab{b}})\citenamefont
  {Guo}, \citenamefont {Vaidya}, \citenamefont {Kroeze}, \citenamefont
  {Lunney}, \citenamefont {Lev},\ and\ \citenamefont
  {Keeling}}]{Guo2019Emergent}%
  \BibitemOpen
  \bibfield  {author} {\bibinfo {author} {\bibfnamefont {Y.}~\bibnamefont
  {Guo}}, \bibinfo {author} {\bibfnamefont {V.~D.}\ \bibnamefont {Vaidya}},
  \bibinfo {author} {\bibfnamefont {R.~M.}\ \bibnamefont {Kroeze}}, \bibinfo
  {author} {\bibfnamefont {R.~A.}\ \bibnamefont {Lunney}}, \bibinfo {author}
  {\bibfnamefont {B.~L.}\ \bibnamefont {Lev}}, \ and\ \bibinfo {author}
  {\bibfnamefont {J.}~\bibnamefont {Keeling}},\ }\bibfield  {title} {\emph
  {\enquote {\bibinfo {title} {{Emergent and broken symmetries of atomic
  self-organization arising from Gouy phase shifts in multimode cavity QED}},}\
  }}\href {\doibase 10.1103/PhysRevA.99.053818} {\bibfield  {journal} {\bibinfo
   {journal} {Phys. Rev. A}\ }\textbf {\bibinfo {volume} {99}},\ \bibinfo
  {pages} {53818} (\bibinfo {year} {2019}{\natexlab{b}})}\BibitemShut {NoStop}%
\bibitem [{\citenamefont {Gopalakrishnan}\ \emph {et~al.}(2012)\citenamefont
  {Gopalakrishnan}, \citenamefont {Lev},\ and\ \citenamefont
  {Goldbart}}]{Gopalakrishnan2012:exploring}%
  \BibitemOpen
  \bibfield  {author} {\bibinfo {author} {\bibfnamefont {S.}~\bibnamefont
  {Gopalakrishnan}}, \bibinfo {author} {\bibfnamefont {B.~L.}\ \bibnamefont
  {Lev}}, \ and\ \bibinfo {author} {\bibfnamefont {P.~M.}\ \bibnamefont
  {Goldbart}},\ }\bibfield  {title} {\emph {\enquote {\bibinfo {title}
  {Exploring models of associative memory via cavity quantum
  electrodynamics},}\ }}\href {\doibase 10.1080/14786435.2011.637980}
  {\bibfield  {journal} {\bibinfo  {journal} {Phil. Mag.}\ }\textbf {\bibinfo
  {volume} {92}},\ \bibinfo {pages} {353} (\bibinfo {year} {2012})}\BibitemShut
  {NoStop}%
\bibitem [{\citenamefont {Fiorelli}\ \emph {et~al.}(2020)\citenamefont
  {Fiorelli}, \citenamefont {Marcuzzi}, \citenamefont {Rotondo}, \citenamefont
  {Carollo},\ and\ \citenamefont {Lesanovsky}}]{Fiorelli2020:Signatures}%
  \BibitemOpen
  \bibfield  {author} {\bibinfo {author} {\bibfnamefont {E.}~\bibnamefont
  {Fiorelli}}, \bibinfo {author} {\bibfnamefont {M.}~\bibnamefont {Marcuzzi}},
  \bibinfo {author} {\bibfnamefont {P.}~\bibnamefont {Rotondo}}, \bibinfo
  {author} {\bibfnamefont {F.}~\bibnamefont {Carollo}}, \ and\ \bibinfo
  {author} {\bibfnamefont {I.}~\bibnamefont {Lesanovsky}},\ }\bibfield  {title}
  {\emph {\enquote {\bibinfo {title} {Signatures of associative memory behavior
  in a multimode dicke model},}\ }}\href {\doibase
  10.1103/PhysRevLett.125.070604} {\bibfield  {journal} {\bibinfo  {journal}
  {Phys. Rev. Lett.}\ }\textbf {\bibinfo {volume} {125}},\ \bibinfo {pages}
  {070604} (\bibinfo {year} {2020})}\BibitemShut {NoStop}%
\bibitem [{\citenamefont {Marsh}\ \emph {et~al.}(2020)\citenamefont {Marsh},
  \citenamefont {Guo}, \citenamefont {Kroeze}, \citenamefont {Gopalakrishnan},
  \citenamefont {Ganguli}, \citenamefont {Keeling},\ and\ \citenamefont
  {Lev}}]{marsh2020enhancing}%
  \BibitemOpen
  \bibfield  {author} {\bibinfo {author} {\bibfnamefont {B.~P.}\ \bibnamefont
  {Marsh}}, \bibinfo {author} {\bibfnamefont {Y.}~\bibnamefont {Guo}}, \bibinfo
  {author} {\bibfnamefont {R.~M.}\ \bibnamefont {Kroeze}}, \bibinfo {author}
  {\bibfnamefont {S.}~\bibnamefont {Gopalakrishnan}}, \bibinfo {author}
  {\bibfnamefont {S.}~\bibnamefont {Ganguli}}, \bibinfo {author} {\bibfnamefont
  {J.}~\bibnamefont {Keeling}}, \ and\ \bibinfo {author} {\bibfnamefont
  {B.~L.}\ \bibnamefont {Lev}},\ }\href@noop {} {\enquote {\bibinfo {title}
  {{Enhancing associative memory recall and storage capacity using confocal
  cavity QED}},}\ } (\bibinfo {year} {2020}),\ \bibinfo {note} {preprint},\
  \Eprint {http://arxiv.org/abs/2009.01227} {2009.01227} \BibitemShut {NoStop}%
\bibitem [{\citenamefont {Minganti}\ \emph {et~al.}(2018)\citenamefont
  {Minganti}, \citenamefont {Biella}, \citenamefont {Bartolo},\ and\
  \citenamefont {Ciuti}}]{Minganti2018:Spectral}%
  \BibitemOpen
  \bibfield  {author} {\bibinfo {author} {\bibfnamefont {F.}~\bibnamefont
  {Minganti}}, \bibinfo {author} {\bibfnamefont {A.}~\bibnamefont {Biella}},
  \bibinfo {author} {\bibfnamefont {N.}~\bibnamefont {Bartolo}}, \ and\
  \bibinfo {author} {\bibfnamefont {C.}~\bibnamefont {Ciuti}},\ }\bibfield
  {title} {\emph {\enquote {\bibinfo {title} {{Spectral theory of Liouvillians
  for dissipative phase transitions}},}\ }}\href {\doibase
  10.1103/PhysRevA.98.042118} {\bibfield  {journal} {\bibinfo  {journal} {Phys.
  Rev. A}\ }\textbf {\bibinfo {volume} {98}},\ \bibinfo {pages} {042118}
  (\bibinfo {year} {2018})}\BibitemShut {NoStop}%
\bibitem [{\citenamefont {Kirton}\ and\ \citenamefont
  {Keeling}(2017)}]{Kirton2017:Suppressing}%
  \BibitemOpen
  \bibfield  {author} {\bibinfo {author} {\bibfnamefont {P.}~\bibnamefont
  {Kirton}}\ and\ \bibinfo {author} {\bibfnamefont {J.}~\bibnamefont
  {Keeling}},\ }\bibfield  {title} {\emph {\enquote {\bibinfo {title}
  {{Suppressing and restoring the dicke superradiance transition by dephasing
  and decay}},}\ }}\href {\doibase 10.1103/PhysRevLett.118.123602} {\bibfield
  {journal} {\bibinfo  {journal} {Phys. Rev. Lett.}\ }\textbf {\bibinfo
  {volume} {118}},\ \bibinfo {pages} {123602} (\bibinfo {year}
  {2017})}\BibitemShut {NoStop}%
\bibitem [{\citenamefont {Keeling}\ \emph {et~al.}(2010)\citenamefont
  {Keeling}, \citenamefont {Bhaseen},\ and\ \citenamefont
  {Simons}}]{Keeling2010:Collective}%
  \BibitemOpen
  \bibfield  {author} {\bibinfo {author} {\bibfnamefont {J.}~\bibnamefont
  {Keeling}}, \bibinfo {author} {\bibfnamefont {M.~J.}\ \bibnamefont
  {Bhaseen}}, \ and\ \bibinfo {author} {\bibfnamefont {B.~D.}\ \bibnamefont
  {Simons}},\ }\bibfield  {title} {\emph {\enquote {\bibinfo {title}
  {{Collective dynamics of Bose-Einstein condensates in optical cavities}},}\
  }}\href {\doibase 10.1103/PhysRevLett.105.043001} {\bibfield  {journal}
  {\bibinfo  {journal} {Phys. Rev. Lett.}\ }\textbf {\bibinfo {volume} {105}},\
  \bibinfo {pages} {043001} (\bibinfo {year} {2010})}\BibitemShut {NoStop}%
\bibitem [{\citenamefont {Palacino}\ and\ \citenamefont {Keeling}(2020)}]{SM}%
  \BibitemOpen
  \bibfield  {author} {\bibinfo {author} {\bibfnamefont {R.}~\bibnamefont
  {Palacino}}\ and\ \bibinfo {author} {\bibfnamefont {J.}~\bibnamefont
  {Keeling}},\ }\href@noop {} {\enquote {\bibinfo {title} {Supplemental
  material},}\ } (\bibinfo {year} {2020}),\ \bibinfo {note} {see supplemental
  material for full expressions of the 4KRE, cumulant equations and comparison
  of Liouvillian gap spacing to the Redfield theory of the Dicke
  model.}\BibitemShut {Stop}%
\bibitem [{\citenamefont {Lindblad}(1976)}]{Lindblad1976b}%
  \BibitemOpen
  \bibfield  {author} {\bibinfo {author} {\bibfnamefont {G.}~\bibnamefont
  {Lindblad}},\ }\bibfield  {title} {\emph {\enquote {\bibinfo {title} {{On the
  generators of quantum dynamical semigroups}},}\ }}\href {\doibase
  10.1007/BF01608499} {\bibfield  {journal} {\bibinfo  {journal} {Commun. Math.
  Phys.}\ }\textbf {\bibinfo {volume} {48}},\ \bibinfo {pages} {119} (\bibinfo
  {year} {1976})}\BibitemShut {NoStop}%
\bibitem [{\citenamefont {Jeske}\ \emph
  {et~al.}(2015{\natexlab{a}})\citenamefont {Jeske}, \citenamefont {Ing},
  \citenamefont {Plenio}, \citenamefont {Huelga},\ and\ \citenamefont
  {Cole}}]{Jeske2014:BlochRedfield}%
  \BibitemOpen
  \bibfield  {author} {\bibinfo {author} {\bibfnamefont {J.}~\bibnamefont
  {Jeske}}, \bibinfo {author} {\bibfnamefont {D.}~\bibnamefont {Ing}}, \bibinfo
  {author} {\bibfnamefont {M.~B.}\ \bibnamefont {Plenio}}, \bibinfo {author}
  {\bibfnamefont {S.~F.}\ \bibnamefont {Huelga}}, \ and\ \bibinfo {author}
  {\bibfnamefont {J.~H.}\ \bibnamefont {Cole}},\ }\bibfield  {title} {\emph
  {\enquote {\bibinfo {title} {{Bloch-Redfield equations for modeling
  light-harvesting complexes}},}\ }}\href {\doibase 10.1063/1.4907370}
  {\bibfield  {journal} {\bibinfo  {journal} {J. Chem. Phys}\ }\textbf
  {\bibinfo {volume} {142}},\ \bibinfo {pages} {064104} (\bibinfo {year}
  {2015}{\natexlab{a}})}\BibitemShut {NoStop}%
\bibitem [{\citenamefont {Eastham}\ \emph {et~al.}(2016)\citenamefont
  {Eastham}, \citenamefont {Kirton}, \citenamefont {Cammack}, \citenamefont
  {Lovett},\ and\ \citenamefont {Keeling}}]{Eastham2016}%
  \BibitemOpen
  \bibfield  {author} {\bibinfo {author} {\bibfnamefont {P.~R.}\ \bibnamefont
  {Eastham}}, \bibinfo {author} {\bibfnamefont {P.}~\bibnamefont {Kirton}},
  \bibinfo {author} {\bibfnamefont {H.~M.}\ \bibnamefont {Cammack}}, \bibinfo
  {author} {\bibfnamefont {B.~W.}\ \bibnamefont {Lovett}}, \ and\ \bibinfo
  {author} {\bibfnamefont {J.}~\bibnamefont {Keeling}},\ }\bibfield  {title}
  {\emph {\enquote {\bibinfo {title} {Bath-induced coherence and the secular
  approximation},}\ }}\href {\doibase 10.1103/PhysRevA.94.012110} {\bibfield
  {journal} {\bibinfo  {journal} {Phys. Rev. A}\ }\textbf {\bibinfo {volume}
  {94}},\ \bibinfo {pages} {012110} (\bibinfo {year} {2016})}\BibitemShut
  {NoStop}%
\bibitem [{\citenamefont {Cammack}\ \emph {et~al.}(2018)\citenamefont
  {Cammack}, \citenamefont {Kirton}, \citenamefont {Stace}, \citenamefont
  {Eastham}, \citenamefont {Keeling},\ and\ \citenamefont
  {Lovett}}]{Cammack2018}%
  \BibitemOpen
  \bibfield  {author} {\bibinfo {author} {\bibfnamefont {H.~M.}\ \bibnamefont
  {Cammack}}, \bibinfo {author} {\bibfnamefont {P.}~\bibnamefont {Kirton}},
  \bibinfo {author} {\bibfnamefont {T.~M.}\ \bibnamefont {Stace}}, \bibinfo
  {author} {\bibfnamefont {P.~R.}\ \bibnamefont {Eastham}}, \bibinfo {author}
  {\bibfnamefont {J.}~\bibnamefont {Keeling}}, \ and\ \bibinfo {author}
  {\bibfnamefont {B.~W.}\ \bibnamefont {Lovett}},\ }\bibfield  {title} {\emph
  {\enquote {\bibinfo {title} {Coherence protection in coupled quantum
  systems},}\ }}\href {\doibase 10.1103/PhysRevA.97.022103} {\bibfield
  {journal} {\bibinfo  {journal} {Phys. Rev. A}\ }\textbf {\bibinfo {volume}
  {97}},\ \bibinfo {pages} {022103} (\bibinfo {year} {2018})}\BibitemShut
  {NoStop}%
\bibitem [{\citenamefont {Dodin}\ \emph {et~al.}(2018)\citenamefont {Dodin},
  \citenamefont {Tscherbul}, \citenamefont {Alicki}, \citenamefont {Vutha},\
  and\ \citenamefont {Brumer}}]{Dodin2018}%
  \BibitemOpen
  \bibfield  {author} {\bibinfo {author} {\bibfnamefont {A.}~\bibnamefont
  {Dodin}}, \bibinfo {author} {\bibfnamefont {T.}~\bibnamefont {Tscherbul}},
  \bibinfo {author} {\bibfnamefont {R.}~\bibnamefont {Alicki}}, \bibinfo
  {author} {\bibfnamefont {A.}~\bibnamefont {Vutha}}, \ and\ \bibinfo {author}
  {\bibfnamefont {P.}~\bibnamefont {Brumer}},\ }\bibfield  {title} {\emph
  {\enquote {\bibinfo {title} {{Secular versus nonsecular Redfield dynamics and
  Fano coherences in incoherent excitation: An experimental proposal}},}\
  }}\href {\doibase 10.1103/PhysRevA.97.013421} {\bibfield  {journal} {\bibinfo
   {journal} {Phys.\ Rev.\ A}\ }\textbf {\bibinfo {volume} {97}},\ \bibinfo
  {pages} {013421} (\bibinfo {year} {2018})}\BibitemShut {NoStop}%
\bibitem [{\citenamefont {Hartmann}\ and\ \citenamefont
  {Strunz}(2020)}]{Hartmann2020:accuracy}%
  \BibitemOpen
  \bibfield  {author} {\bibinfo {author} {\bibfnamefont {R.}~\bibnamefont
  {Hartmann}}\ and\ \bibinfo {author} {\bibfnamefont {W.~T.}\ \bibnamefont
  {Strunz}},\ }\bibfield  {title} {\emph {\enquote {\bibinfo {title} {Accuracy
  assessment of perturbative master equations: Embracing nonpositivity},}\
  }}\href {\doibase 10.1103/PhysRevA.101.012103} {\bibfield  {journal}
  {\bibinfo  {journal} {Phys. Rev. A}\ }\textbf {\bibinfo {volume} {101}},\
  \bibinfo {pages} {012103} (\bibinfo {year} {2020})}\BibitemShut {NoStop}%
\bibitem [{\citenamefont {D{\"{u}}mcke}\ and\ \citenamefont
  {Spohn}(1979)}]{Duemcke1979:Proper}%
  \BibitemOpen
  \bibfield  {author} {\bibinfo {author} {\bibfnamefont {R.}~\bibnamefont
  {D{\"{u}}mcke}}\ and\ \bibinfo {author} {\bibfnamefont {H.}~\bibnamefont
  {Spohn}},\ }\bibfield  {title} {\emph {\enquote {\bibinfo {title} {{The
  proper form of the generator in the weak coupling limit}},}\ }}\href
  {\doibase 10.1007/BF01325208} {\bibfield  {journal} {\bibinfo  {journal} {Z.
  Physik B}\ }\textbf {\bibinfo {volume} {34}},\ \bibinfo {pages} {419}
  (\bibinfo {year} {1979})}\BibitemShut {NoStop}%
\bibitem [{\citenamefont {Gardiner}(2009)}]{gardiner2009stochastic}%
  \BibitemOpen
  \bibfield  {author} {\bibinfo {author} {\bibfnamefont {C.}~\bibnamefont
  {Gardiner}},\ }\href@noop {} {\emph {\bibinfo {title} {Stochastic methods}}}\
  (\bibinfo  {publisher} {Springer},\ \bibinfo {address} {Berlin},\ \bibinfo
  {year} {2009})\BibitemShut {NoStop}%
\bibitem [{Note1()}]{Note1}%
  \BibitemOpen
  \bibinfo {note} {We label eigenvalues as: $\lambda ^{(k)}_{i}$, with
  $i=0,1,2,\protect \ldots $. Hence, $\lambda ^{(0)}_{0}=0$ is the zero-mode
  associated to the steady state and $\lambda ^{(0)}_{1}$ is the first non-zero
  eigenvalue in the $k=0$ sector.}\BibitemShut {Stop}%
\bibitem [{\citenamefont {Bu{\v{c}}a}\ and\ \citenamefont
  {Jaksch}(2019)}]{Buca2019:Dissipation}%
  \BibitemOpen
  \bibfield  {author} {\bibinfo {author} {\bibfnamefont {B.}~\bibnamefont
  {Bu{\v{c}}a}}\ and\ \bibinfo {author} {\bibfnamefont {D.}~\bibnamefont
  {Jaksch}},\ }\bibfield  {title} {\emph {\enquote {\bibinfo {title}
  {Dissipation induced nonstationarity in a quantum gas},}\ }}\href {\doibase
  10.1103/PhysRevLett.123.260401} {\bibfield  {journal} {\bibinfo  {journal}
  {Phys. Rev. Lett.}\ }\textbf {\bibinfo {volume} {123}},\ \bibinfo {pages}
  {260401} (\bibinfo {year} {2019})}\BibitemShut {NoStop}%
\bibitem [{\citenamefont {Chiacchio}\ and\ \citenamefont
  {Nunnenkamp}(2019)}]{Chiacchio2019:Dissipation}%
  \BibitemOpen
  \bibfield  {author} {\bibinfo {author} {\bibfnamefont {E.~I.~R.}\
  \bibnamefont {Chiacchio}}\ and\ \bibinfo {author} {\bibfnamefont
  {A.}~\bibnamefont {Nunnenkamp}},\ }\bibfield  {title} {\emph {\enquote
  {\bibinfo {title} {Dissipation-induced instabilities of a spinor
  bose-einstein condensate inside an optical cavity},}\ }}\href {\doibase
  10.1103/PhysRevLett.122.193605} {\bibfield  {journal} {\bibinfo  {journal}
  {Phys. Rev. Lett.}\ }\textbf {\bibinfo {volume} {122}},\ \bibinfo {pages}
  {193605} (\bibinfo {year} {2019})}\BibitemShut {NoStop}%
\bibitem [{\citenamefont {Jeske}\ \emph
  {et~al.}(2015{\natexlab{b}})\citenamefont {Jeske}, \citenamefont {Ing},
  \citenamefont {Plenio}, \citenamefont {Huelga},\ and\ \citenamefont
  {Cole}}]{Jeske2014a}%
  \BibitemOpen
  \bibfield  {author} {\bibinfo {author} {\bibfnamefont {J.}~\bibnamefont
  {Jeske}}, \bibinfo {author} {\bibfnamefont {D.}~\bibnamefont {Ing}}, \bibinfo
  {author} {\bibfnamefont {M.~B.}\ \bibnamefont {Plenio}}, \bibinfo {author}
  {\bibfnamefont {S.~F.}\ \bibnamefont {Huelga}}, \ and\ \bibinfo {author}
  {\bibfnamefont {J.~H.}\ \bibnamefont {Cole}},\ }\bibfield  {title} {\emph
  {\enquote {\bibinfo {title} {{Bloch-Redfield equations for modeling
  light-harvesting complexes}},}\ }}\href {\doibase 10.1063/1.4907370}
  {\bibfield  {journal} {\bibinfo  {journal} {J. Chem. Phys}\ }\textbf
  {\bibinfo {volume} {142}},\ \bibinfo {pages} {064104} (\bibinfo {year}
  {2015}{\natexlab{b}})}\BibitemShut {NoStop}%
\end{thebibliography}
\end{document}